\documentclass[10pt,journal,compsoc]{IEEEtran}
\ifCLASSOPTIONcompsoc
  \usepackage[nocompress]{cite}
\else
  \usepackage{cite}
\fi
\ifCLASSINFOpdf
\else
\fi
\usepackage{color}
\usepackage{cite}
\usepackage{booktabs} % For formal tables
\usepackage{multirow}
\usepackage{subfigure}
\usepackage{graphicx}
\usepackage{bm}
\usepackage{url}
\usepackage{amsmath}
\usepackage{amssymb}
\usepackage{amsfonts}
\usepackage{mathrsfs}
\usepackage{rotating}
\usepackage{footmisc}
\usepackage{footnote}
\usepackage{scrextend}
\usepackage{hyperref}
\usepackage{float}
\newtheorem{definition}{Definition}
\makesavenoteenv{tabular}
\newcommand{\tabincell}[2]{\linespread{0.8}\selectfont\begin{tabular}{@{}#1@{}}#2\end{tabular}}
\begin{document}
%
% paper title
% Titles are generally capitalized except for words such as a, an, and, as,
% at, but, by, for, in, nor, of, on, or, the, to and up, which are usually
% not capitalized unless they are the first or last word of the title.
% Linebreaks \\ can be used within to get better formatting as desired.
% Do not put math or special symbols in the title.
\title{Network Representation Learning: A Survey}
%
%
% author names and IEEE memberships
% note positions of commas and nonbreaking spaces ( ~ ) LaTeX will not break
% a structure at a ~ so this keeps an author's name from being broken across
% two lines.
% use \thanks{} to gain access to the first footnote area
% a separate \thanks must be used for each paragraph as LaTeX2e's \thanks
% was not built to handle multiple paragraphs
%
%
%\IEEEcompsocitemizethanks is a special \thanks that produces the bulleted
% lists the Computer Society journals use for "first footnote" author
% affiliations. Use \IEEEcompsocthanksitem which works much like \item
% for each affiliation group. When not in compsoc mode,
% \IEEEcompsocitemizethanks becomes like \thanks and
% \IEEEcompsocthanksitem becomes a line break with idention. This
% facilitates dual compilation, although admittedly the differences in the
% desired content of \author between the different types of papers makes a
% one-size-fits-all approach a daunting prospect. For instance, compsoc 
% journal papers have the author affiliations above the "Manuscript
% received ..."  text while in non-compsoc journals this is reversed. Sigh.

\author{Daokun Zhang,
	Jie Yin,
	Xingquan Zhu \IEEEmembership{Senior Member,~IEEE},
	Chengqi Zhang \IEEEmembership{Senior Member,~IEEE}
	\IEEEcompsocitemizethanks{\IEEEcompsocthanksitem Daokun Zhang and Chengqi Zhang are with the Centre for Artificial Intelligence, FEIT, University of Technology Sydney, Australia
	\protect\\
	Email: Daokun.Zhang@student.uts.edu.au, Chengqi.Zhang@uts.edu.au.
	\IEEEcompsocthanksitem Jie Yin is with the Discipline of Business Analytics, The University of Sydney, Australia.
	\protect\\
	Email: jie.yin@sydney.edu.au.
	\IEEEcompsocthanksitem Xingquan Zhu is with the Dept. of CEECS, Florida Atlantic University, USA.
	\protect\\
	Email: xqzhu@cse.fau.edu.}% <-this % stops an unwanted space
\thanks{Manuscript received 3 Dec., 2017; revised 26 Apr. 2018; accepted 12 June 2018. (Corresponding author: Jie Yin.)}}

\IEEEtitleabstractindextext{%
\begin{abstract}
With the widespread use of information technologies, information networks are becoming increasingly popular to capture complex relationships across various disciplines, such as social networks, citation networks, telecommunication networks, and biological networks. Analyzing these networks sheds light on different aspects of social life such as the structure of societies, information diffusion, and communication patterns. In reality, however, the large scale of information networks often makes network analytic tasks computationally expensive or intractable. Network representation learning has been recently proposed as a new learning paradigm to embed network vertices into a low-dimensional vector space, by preserving network topology structure, vertex content, and other side information. This facilitates the original network to be easily handled in the new vector space for further analysis. In this survey, we perform a comprehensive review of the current literature on network representation learning in the data mining and machine learning field. We propose new taxonomies to categorize and summarize the state-of-the-art network representation learning techniques according to the underlying learning mechanisms, the network information intended to preserve, as well as the algorithmic designs and methodologies. We summarize evaluation protocols used for validating network representation learning including published benchmark datasets, evaluation methods, and open source algorithms. We also perform empirical studies to compare the performance of representative algorithms on common datasets, and analyze their computational complexity. Finally, we suggest promising research directions to facilitate future study.
\end{abstract}

% Note that keywords are not normally used for peerreview papers.
\begin{IEEEkeywords}
Information networks, graph mining, network representation learning, network embedding.
\end{IEEEkeywords}}

% make the title area
\maketitle

% To allow for easy dual compilation without having to reenter the
% abstract/keywords data, the \IEEEtitleabstractindextext text will
% not be used in maketitle, but will appear (i.e., to be "transported")
% here as \IEEEdisplaynontitleabstractindextext when the compsoc 
% or transmag modes are not selected <OR> if conference mode is selected 
% - because all conference papers position the abstract like regular
% papers do.
\IEEEdisplaynontitleabstractindextext
% \IEEEdisplaynontitleabstractindextext has no effect when using
% compsoc or transmag under a non-conference mode.

% For peer review papers, you can put extra information on the cover
% page as needed:
% \ifCLASSOPTIONpeerreview
% \begin{center} \bfseries EDICS Category: 3-BBND \end{center}
% \fi
%
% For peerreview papers, this IEEEtran command inserts a page break and
% creates the second title. It will be ignored for other modes.
\IEEEpeerreviewmaketitle

\IEEEraisesectionheading{\section{Introduction}\label{sec:introduction}}
% Computer Society journal (but not conference!) papers do something unusual
% with the very first section heading (almost always called "Introduction").
% They place it ABOVE the main text! IEEEtran.cls does not automatically do
% this for you, but you can achieve this effect with the provided
% \IEEEraisesectionheading{} command. Note the need to keep any \label that
% is to refer to the section immediately after \section in the above as
% \IEEEraisesectionheading puts \section within a raised box.

% The very first letter is a 2 line initial drop letter followed
% by the rest of the first word in caps (small caps for compsoc).
% 
% form to use if the first word consists of a single letter:
% \IEEEPARstart{A}{demo} file is ....
% 
% form to use if you need the single drop letter followed by
% normal text (unknown if ever used by the IEEE):
% \IEEEPARstart{A}{}demo file is ....
% 
% Some journals put the first two words in caps:
% \IEEEPARstart{T}{his demo} file is ....
% 
% Here we have the typical use of a "T" for an initial drop letter
% and "HIS" in caps to complete the first word.
\IEEEPARstart{I}nformation networks are becoming ubiquitous across a large spectrum of real-world applications in forms of social networks, citation networks, telecommunication networks and biological networks, \textit{etc.} The scale of these networks ranges from hundreds to millions or even billions of vertices~\cite{tang2015line}. Analyzing information networks plays a crucial role in a variety of emerging applications across many disciplines. For example, in social networks, classifying users into meaningful social groups is useful for many important tasks, such as user search, targeted advertising and recommendations; in communication networks, detecting community structures can help better understand the rumor spreading process; in biological networks, inferring interactions between proteins can facilitate new treatments for diseases. Nevertheless, efficient analysis of these networks heavily relies on the ways how networks are represented. Often, a discrete adjacency matrix is used to represent a network, which only captures neighboring relationships between vertices. Indeed, this simple representation cannot embody more complex, higher-order structure relationships, such as paths, frequent substructure \textit{etc.} As a result, such a traditional routine often makes many network analytic tasks computationally expensive and intractable over large-scale networks. Taking community detection as an example, most existing algorithms involve calculating the spectral decomposition of a matrix~\cite{malliaros2013clustering} with at least quadratic time complexity with respect to the number of vertices. This computational overhead makes algorithms hard to scale to large-scale networks with millions of vertices.

Recently, network representation learning (NRL) has aroused a lot of research interest. NRL aims to learn latent, low-dimensional representations of network vertices, while preserving network topology structure, vertex content, and other side information. After new vertex representations are learned, network analytic tasks can be easily and efficiently carried out by applying conventional vector-based machine learning algorithms to the new representation space. This obviates the necessity for deriving complex algorithms that are applied directly on the original network.

Earlier work related to network representation learning dates back to the early 2000s, when researchers proposed graph embedding algorithms as part of dimensionality reduction techniques. Given a set of \textit{i.i.d.} (independent and identically distributed) data points as input, graph embedding algorithms first calculate the similarity between pairwise data points to construct an affinity graph, \textit{e.g.}, the $k$-nearest neighbor graph, and then embed the affinity graph into a new space having much lower dimensionality. The idea is to find a low-dimensional manifold structure hidden in the high-dimensional data geometry reflected by the constructed graph, so that connected vertices are kept closer to each other in the new embedding space. Isomap~\cite{tenenbaum2000global}, Locally Linear Embedding (LLE)~\cite{roweis2000nonlinear} and Laplacian Eigenmap~\cite{belkin2002laplacian} are examples of algorithms based on this rationale. However, graph embedding algorithms are designed on \textit{i.i.d.} data mainly for dimensionality reduction purpose. Most of these algorithms usually have at least quadratic time complexity with respect to the number of vertices, so the scalability is a major issue when they are applied to large-scale networks.

Since 2008, significant research efforts have shifted to the development of effective and scalable representation learning techniques that are directly designed for complex information networks. Many NRL algorithms, \textit{e.g.}, ~\cite{perozzi2014deepwalk,yang2015network,zhang2016collective,cao2016deep}, have been proposed to embed existing networks, showing promising performance for various applications. These algorithms embed a network into a latent, low-dimensional space that preserves structure proximity and attribute affinity, such that the original vertices of the network can be represented as low-dimensional vectors. The resulting compact, low-dimensional vector representations can be then taken as features to any vector-based machine learning algorithms. This paves the way for a wide range of network analytic tasks to be easily and efficiently tackled in the new vector space, such as node classification~\cite{zhu2007combining,bhagat2011node}, link prediction~\cite{lu2011link,gao2011temporal}, clustering~\cite{malliaros2013clustering}, recommendation~\cite{zhang2017regions,xie2016learning}, similarity search~\cite{liu2018distance}, and visualization~\cite{tang2016visualizing}. Using vector representation to represent complex networks has now been gradually advanced to many other domains, such as point-of-interest recommendation in urban computing~\cite{xie2016learning}, and knowledge graph search~\cite{lin2015learning} in knowledge engineering and database systems.

\subsection{Challenges}

Despite its great potential, network representation learning is inherently difficult and is confronted with several key challenges that we summarize as follows.

\noindent\textbf{Structure-preserving}: To learn informative vertex representations, network representation learning should preserve network structure, such that vertices similar/close to each other in the original structure space should also be represented similarly in the learned vector space. However, as stated in \cite{wang2016structural,zhang2016homophily}, the structure-level similarity between vertices is reflected not only at the local neighborhood structure but also at the more global community structure. Therefore, the local and global structure should be simultaneously preserved in network representation learning. 	
	
\noindent\textbf{Content-preserving}: Besides structure information, vertices of many networks are attached with rich content on attributes. Vertex attributes not only exert huge impacts on the forming of networks, but also provide direct evidence to measure attribute-level similarity between vertices. Therefore, if properly imported, attribute content can compensate network structure to render more informative vertex representations. However, due to heterogeneity of the two information sources, how to effectively leverage vertex attributes and make them compensate rather than deteriorate network structure is an open research problem.
	
\noindent\textbf{Data sparsity}: For many real-world information networks, due to the privacy or legal restrictions, the problem of data sparsity exists in both network structure and vertex content. At the structure level, only very limited links are sometimes observed, making it difficult to discover the structure-level relatedness between vertices that are not explicitly connected. At the vertex content level, many values of vertex attributes are usually missing, which increases the difficulty of measuring content-level vertex similarity. Thus, it is challenging for network representation learning to overcome the data sparsity problem.
	
\noindent\textbf{Scalability}: Real-world networks, social networks in particular, consist of millions or billions of vertices. The large scale of the networks challenges not only the traditional network analytic tasks but also the newborn network representation learning task. Without special concern, learning vertex representations for large-scale networks with limited computing resources may cost months of time, which is practically infeasible, especially for the case involving a large number of trails for tuning parameters. Therefore, it is necessary to design NRL algorithms that can learn vertex representations efficiently and meanwhile guarantee the effectiveness for large-scale networks.

\subsection{Our Contribution}
This survey provides a comprehensive up-to-date review of the state-of-the-art network representation learning techniques, with a focus on the learning of vertex representations. It covers not only early work on preserving network structure, but also a new surge of recent studies that incorporate vertex content and/or vertex labels as auxiliary information into the learning process of network embedding. By doing so, we hope to provide a useful guideline for the research community to better understand (1) new taxonomies of network representation learning methods, (2) the characteristics, uniqueness, and the niche of different types of network embedding methods, and (3) the resources and future challenges to stimulate research in the area. In particular, this survey has four major contributions:
\begin{itemize}
	
	\item We propose new taxonomies to categorize existing network representation learning techniques according to the underlying learning mechanisms, the network information intended to preserve, as well as the algorithmic designs and methodologies. As a result, this survey provides new angles to better understand the existing work.
	
	\item We provide a detailed and thorough study of the state-of-the-art network representation learning algorithms. Compared to the existing graph embedding surveys, we not only review a more comprehensive set of research work on network representation learning, but also provide multifaceted algorithmic perspectives to understand the advantages and disadvantages of different algorithms.
	
	\item We summarize evaluation protocols used for validating network representation learning techniques, including published benchmark datasets, evaluation methods, and open source algorithms. We also perform empirical studies to compare the performance of representative algorithms, along with a detailed analysis of computational complexity.
	
	\item To foster future research, we suggest six promising future research directions for network representation learning, and summarize the limitations of current research work and propose new research ideas for each direction. 
	
\end{itemize}

\subsection{Related Surveys and Differences}

A few graph embedding and representation learning related surveys exist in the recent literature. The first is \cite{moyano2017learning}, which reviews a few representative methods for network representation learning and visits some key concepts around the idea of representation learning and its connections to other related field such as dimensionality reduction, deep learning, and network science. \cite{palash2017graph} categorizes representative network embedding algorithms from a methodology perspective. \cite{hamilton2017representation} reviews a few representation learning methods for embedding individual vertices as well as subgraphs, especially those inspired by deep learning, within an encoder-decoder framework. Yet, the majority of embedding algorithms reviewed by these surveys primarily preserve network structure. Recently, \cite{cui2017survey,cai2018tkde} extend to cover work leveraging other side information, such as vertex attributes and/or vertex labels, to harness representation learning. 

In summary, existing surveys have the following limitations. First, they typically focus on one single taxonomy to categorize the existing work. None of them provides a multifaceted view to analyze the state-of-the-art network representation learning techniques and to compare their advantages and disadvantages. Second, existing surveys do not have in-depth analysis of algorithm complexity and optimization methods, or they do not provide empirical results to compare the performance of different algorithms. Third, there is a lack of summary on available resources, such as publicly available datasets and open source algorithms, to facilitate future research. In this work, we provide the most comprehensive survey to bridge the gap. We believe that this survey will benefit both researchers and practitioners to gain a deep understanding of different approaches, and provide rich resources to foster future research in the field.

%Different from previous surveys, we conduct a review of the most recent literature to cover the state-of-the-art network representation learning techniques. We provide a systematic categorization and detailed analysis of the existing techniques from two new algorithmic perspectives: the information sources and the methodology. This makes it easier to understand the strength and weakness of different algorithms. We summarize evaluation protocols used for validating different algorithms, including dataset resources, evaluation methods, as well as current open-source algorithms. More importantly, we report empirical results to compare the performance of representative NRL algorithms on common network analytical tasks, along with an in-depth analysis of their computational complexity. We believe this survey will benefit both researchers and practitioners to gain a deep understand of different approaches, and provide rich resources to foster future research in the field.

\subsection{Organization of the Survey}
The rest of this survey is organized as follows. In Section~\ref{sec:prelim}, we provide preliminaries and definitions required to understand the problem and the models discussed next. Section~\ref{sec:taxonomy} proposes new taxonomies to categorize the existing network representation learning techniques. Section~\ref{sec:unsupervised} and Section~\ref{sec:semi} review representative algorithms in two categories, respectively. A list of successful applications of network representation learning are discussed in Section~\ref{sec:app}. In Section~\ref{sec:eva}, we summarize the evaluation protocols used to validate network representation learning, along with a comparison of algorithm performance and complexity. We discuss potential research directions in Section~\ref{sec:research-direction}, and conclude the survey in Section~\ref{sec:conclusion}.

\section{Notations and Definitions}\label{sec:prelim}

In this section, as preliminaries, we first define important terminologies that are used to discuss the models next, followed by a formal definition of the network representation learning problem. For ease of presentation, we first define a list of common notations that will be used throughout the survey, as shown in Table ~\ref{table-notation}.

\begin{table}[h]
	\scriptsize
	\renewcommand\arraystretch{1.2}
	\caption{A summary of common notations}
	\centering
	\begin{tabular}{|c|c|}
		\hline
		$G$ & The given information network\\\hline
		$V$ & Set of vertices in the given information network\\\hline
		$E$ & Set of edges in the given information network\\\hline
		$|V|$ & Number of vertices\\\hline
		$|E|$ & Number of edges\\\hline
		$m$ & Number of vertex attributes\\\hline
		$d$ & Dimension of learned vertex representations\\\hline
		$X\in \mathbb{R}^{|V|\times m}$ & The vertex attribute matrix\\\hline
		$\mathcal{Y}$ & Set of vertex labels\\\hline
		$|\mathcal{Y}|$ & Number of vertex labels\\\hline
		$Y\in\mathbb{R}^{|V|\times |\mathcal{Y}|}$ & The vertex label matrix \\\hline
	\end{tabular}
	\label{table-notation}
\end{table}

\begin{definition}[Information Network] An information network is defined as $G=(V,E,X,Y)$, where $V$ denotes a set of vertices, and $|V|$ denotes the number of vertices in network $G$. $E\subseteq(V\times V)$ denotes a set of edges connecting the vertices. $X\in\mathbb{R}^{|V|\times m}$ is the vertex attribute matrix, where $m$ is the number of attributes, and the element $X_{ij}$ is the value of the $i$-th vertex on the $j$-th attribute. $Y\in\mathbb{R}^{|V|\times|\mathcal{Y}|}$ is the vertex label matrix with $\mathcal{Y}$ being a set of labels. If the $i$-th vertex has the $k$-th label, the element $Y_{ik}=1$; otherwise, $Y_{ik}=-1$. Due to privacy concern or information access difficulty, vertex attribute matrix $X$ is often sparse and vertex label matrix $Y$ is usually unobserved or partially observed. For each $(v_{i},v_{j})\in E$, if information network $G$ is undirected, we have $(v_{j},v_{i})\in E$; if $G$ is directed, $(v_{j},v_{i})$ unnecessarily belongs to $E$.\footnote{Without any specific declaration, the networks discussed in this survey are assumed to be undirected.} Each edge $(v_{i},v_{j})\in E$ is also associated to a weight $w_{ij}$, which is equal to $1$, if the information network is binary (unweighted).
\end{definition}

Intuitively, the generation of information networks is not groundless, but guided or dominated by certain latent mechanisms. Although the latent mechanisms are hardly known, they can be reflected by some network properties that widely exist in information networks. Hence, the common network properties are essential for the learning of vertex representations that are informative to accurately interpret information networks. Below, we introduce several common network properties.

\begin{definition}[First-order Proximity]The first-order proximity is the local pairwise proximity between two connected vertices~\cite{tang2015line}. For each vertex pair $(v_{i},v_{j})$, if $(v_{i},v_{j})\in E$, the first-order proximity between $v_{i}$ and $v_{j}$ is $w_{ij}$; otherwise, the first-order proximity between $v_{i}$ and $v_{j}$ is 0. The first-order proximity captures the direct neighbor relationships between vertices.
	
	%Due to its locality, the first-order proximity is a kind of microscopic structure property.
\end{definition}

\begin{definition}[Second-order Proximity and High-order Proximity]The second-order proximity captures the $2$-step relations between each pair of vertices~\cite{tang2015line}. For each vertex pair $(v_{i},v_{j})$, the second order proximity is determined by the number of common neighbors shared by the two vertices, which can also be measured by the $2$-step transition probability from $v_{i}$ to $v_{j}$ equivalently. Compared with the second-order proximity, the high-order proximity \cite{cao2015grarep} captures more global structure, which explores $k$-step $(k\geq 3)$ relations between each pair of vertices. For each vertex pair $(v_{i},v_{j})$, the higher-order proximity is measured by the $k$-step $(k\geq 3)$ transition probability from vertex $v_{i}$ to vertex $v_{j}$, which can also be reflected by the number of $k$-step $(k\geq 3)$ paths from $v_{i}$ to $v_{j}$. The second-order and high-order proximity capture the similarity between a pair of, indirectly connected, vertices with similar structural contexts.
	
	%Though the second-order proximity and the high-order proximity reflects more global structure information than the first-order proximity, they still belongs to microscopic structure property.
\end{definition}

%\begin{definition}[Local Closeness Proximity] The first-order, second-order and high-order proximity describes the proximity between vertices that are close to each in their local neighborhood structure. We define the first-order, second-order and high-order proximity as the local closeness proximity.
%\end{definition}

\begin{figure}[tb]
	\centering
	\scalebox{0.35}{
		\includegraphics[width=\textwidth]{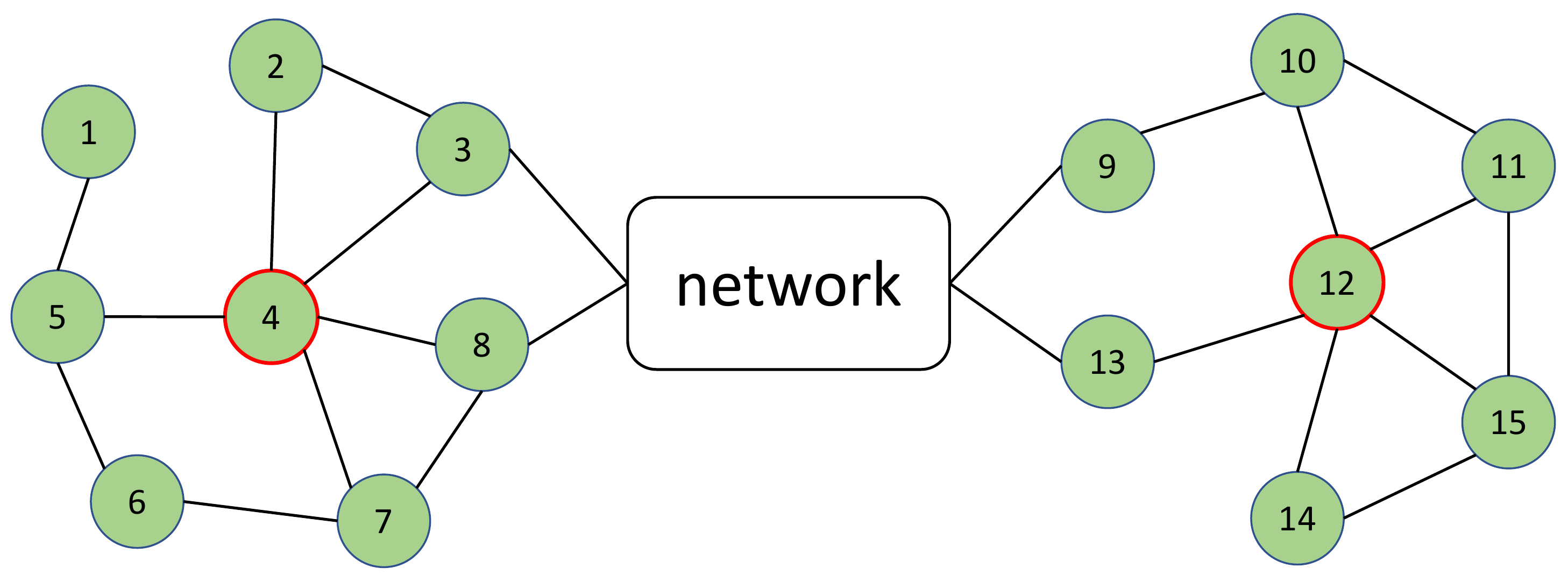}}
	\caption{An illustrative example of structural role proximity. Vertex 4 and vertex 12 have similar structural roles, but are located far away from each other.}
	\label{fig:structuralRole}
\end{figure}

\begin{definition}[Structural Role Proximity]The structural role proximity depicts similarity between vertices serving as the similar roles in their neighborhood, such as edge of a chain, center of a star, and a bridge between two communities. In communication and traffic networks, vertices' structural roles are important to characterize their properties. Different from the first-order, second-order and high-order proximity, which capture the similarity between vertices close to each other in the network, the structural role proximity tries to discover the similarity between distant vertices while sharing the equivalent structural roles.  As is shown in Fig. \ref{fig:structuralRole}, vertex 4 and vertex 12 are located far away from each other, while they serve as the same structural role, center of a star. Thus, they have high structural role proximity.
\end{definition}

\begin{definition}[Intra-community Proximity]The intra-community proximity is the pairwise proximity between vertices in a same community. Many networks have community structure, where vertex-vertex connections within the same community are dense, but connections to vertices outside the community are sparse~\cite{girvan2002community}. As cluster structure, a community preserves certain kinds of common properties of vertices within it. For example, in social networks, communities might represent social groups by interest or background; in citation networks, communities might represent related papers on a same topic. The intra-community proximity captures such cluster structure by preserving the common property shared by vertices within a same community~\cite{wang2017community}.
\end{definition}

%The first-order, second-order and high-order proximity describes the proximity between vertices that are close to each in their local neighborhood structure. We define the first-order, second-order and high-order proximity as \textbf{microscopic structure}. The structural role proximity reflects the similarity between vertices in terms of their local structural roles, and the intra-community proximity captures the community structure. These two kinds of proximity measures the vertex similarity in a mesoscopic view and we categorize them into \textbf{mesoscopic structure}. In addition, networks can also be characterized by certain global properties, such as the scale-free property and small world property, which we define as \textbf{macroscopic structure}.

\textbf{Vertex attribute}: In addition to network structure, vertex attributes can provide direct evidence to measure content-level similarity between vertices. As shown in~\cite{yang2015network,wang2016paired,zhang2016homophily}, vertex attributes and network structure can help each other filter out noisy information and compensate each other to jointly learn informative vertex representations. 

\textbf{Vertex label}: Vertex labels provide direct information about the semantic categorization of each network vertex to certain classes or groups. Vertex labels are strongly influenced by and inherently correlated to both network structure and vertex attributes~\cite{huang2017label}. Though vertex labels are usually partially observed, when coupled with network structure and vertex attributes, they encourage a network structure and vertex attribute consistent labeling, and help learn informative and discriminative vertex representations.

\begin{definition}[Network Representation Learning (NRL)] Given an information network $G=(V,E,X,Y)$, by integrating network structure in $E$, vertex attributes in $X$ and vertex labels in $Y$ (if available), the task of network representation learning is to learn a mapping function $f:v\longmapsto r_{v}\in\mathbb{R}^{d}$, where $r_{v}$ is the learned vector representation of vertex $v$, and $d$ is the dimension of the learned representation. The transformation $f$ preserves the original network information, such that two vertices similar in the original network should also be represented similarly in the learned vector space.
\end{definition}	

The learned vertex representations should satisfy the following conditions: (1) \textit{low-dimensional},\textit{i.e.}, $d\ll |V|$, in other words, the dimension of learned vertex representations should be much smaller than the dimension of the original adjacency matrix representation for memory efficiency and the scalability of subsequent network analytic tasks; (2) \textit{informative}, \textit{i.e.}, the learned vertex representations should preserve vertex proximity reflected by network structure, vertex attributes and vertex labels (if available); (3) \textit{continuous}, \textit{i.e.}, the learned vertex representations should have continuous real values to support subsequent network analytic tasks, like vertex classification, vertex clustering, or anomaly detection, and have smooth decision boundaries to ensure the robustness of these tasks.

\begin{figure}[!htbp]
	\centering
	\subfigure[{\scriptsize Input: Information Network}]{
		\label{fig:NRL:subfig:network}
		\includegraphics[width=1.6in]{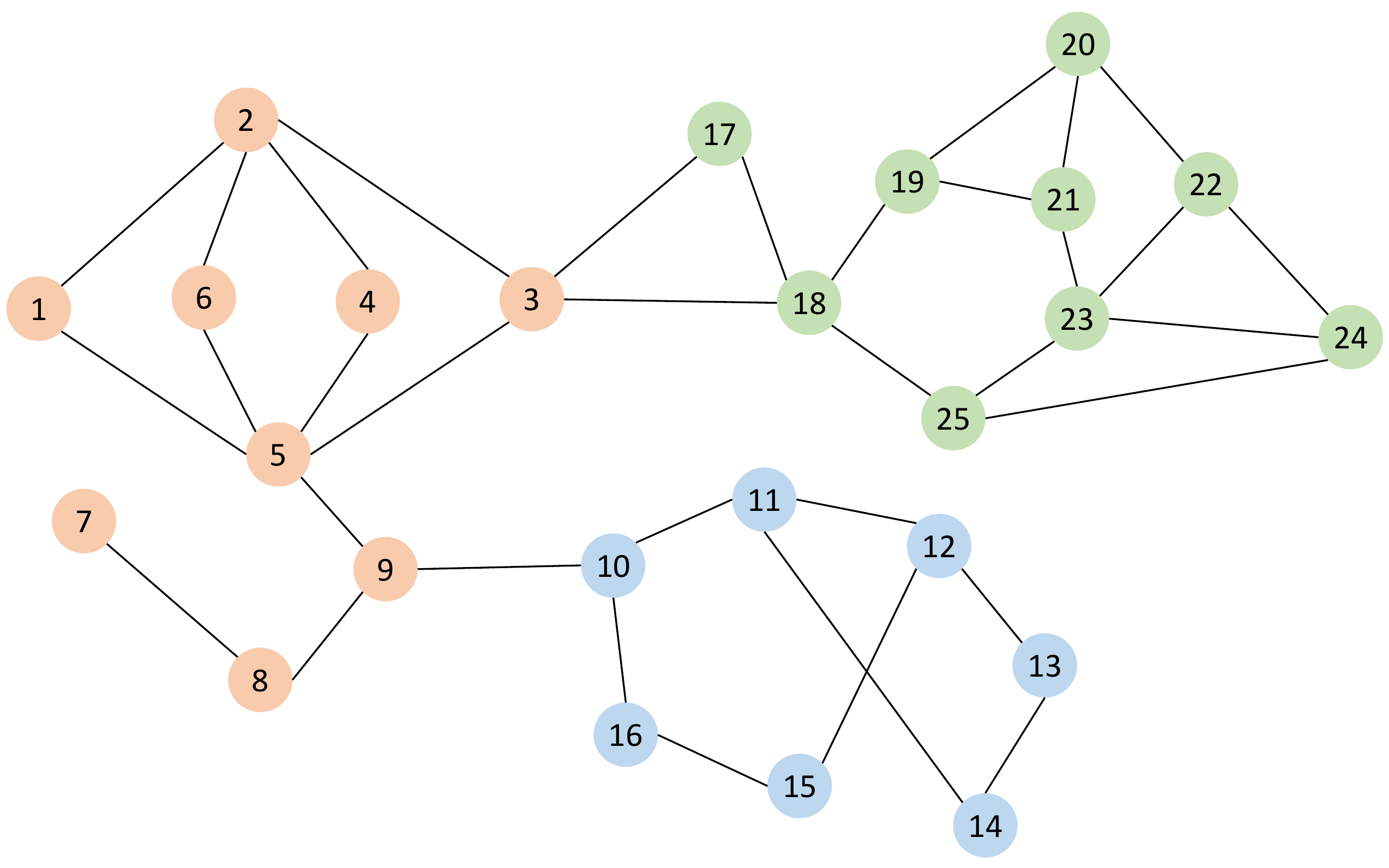}}
	\subfigure[{\scriptsize Output: Vertex Representations}]{
		\label{fig:NRL:subfig:embedding}
		\includegraphics[width=1.6in]{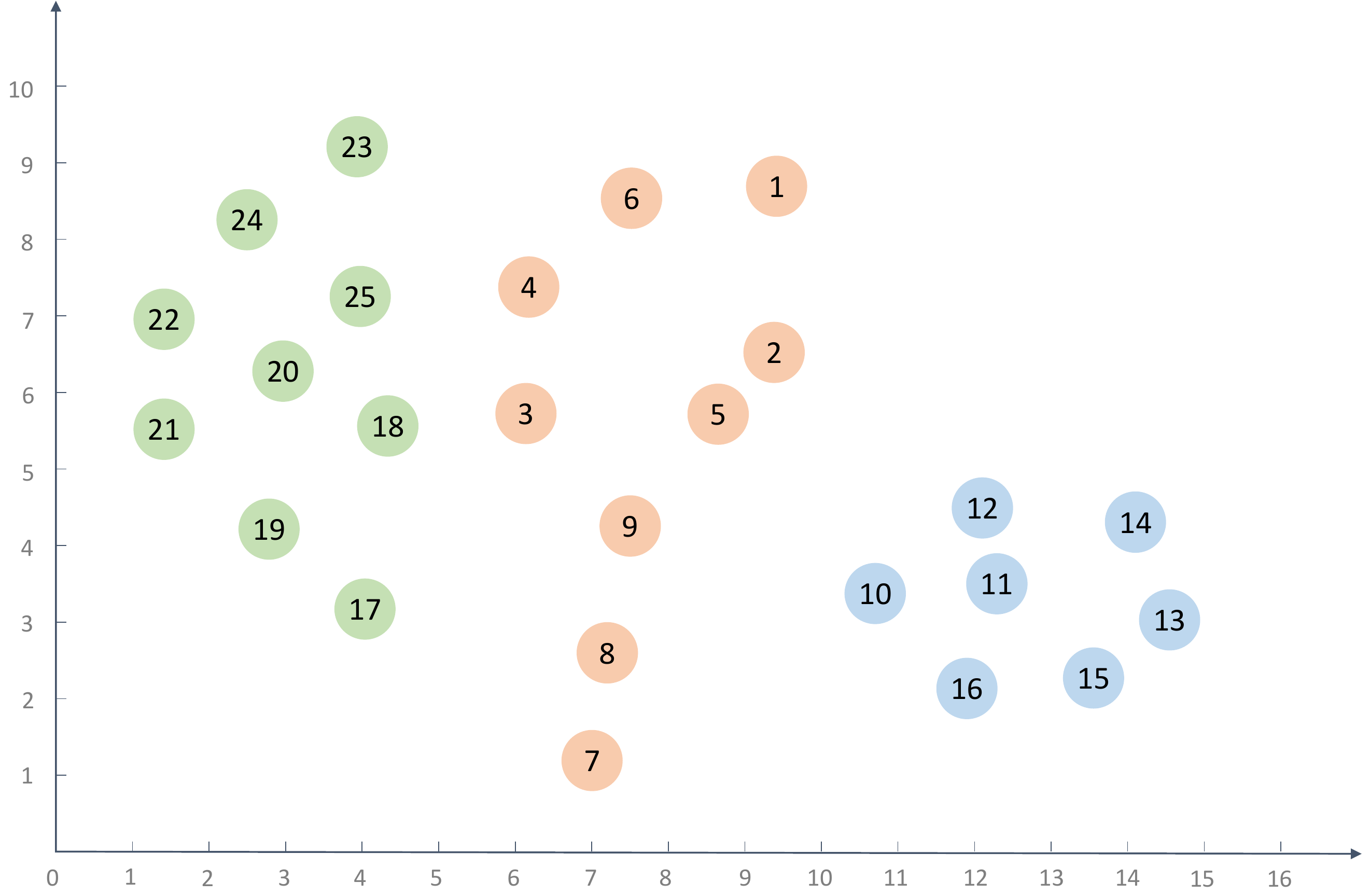}}
	\caption{A conceptual view of network representation learning. Vertices in (a) are indexed using their ID and color coded based on their community information. The network representation learning in (b) transforms all vertices into a two-dimensional vector space, such that vertices with structural proximity are close to each other in the new embedding space. }
	\label{fig:NRL}
\end{figure}

\begin{figure*}[!htbp]
	\centering
	\scalebox{0.75}{
		\includegraphics[width=\textwidth]{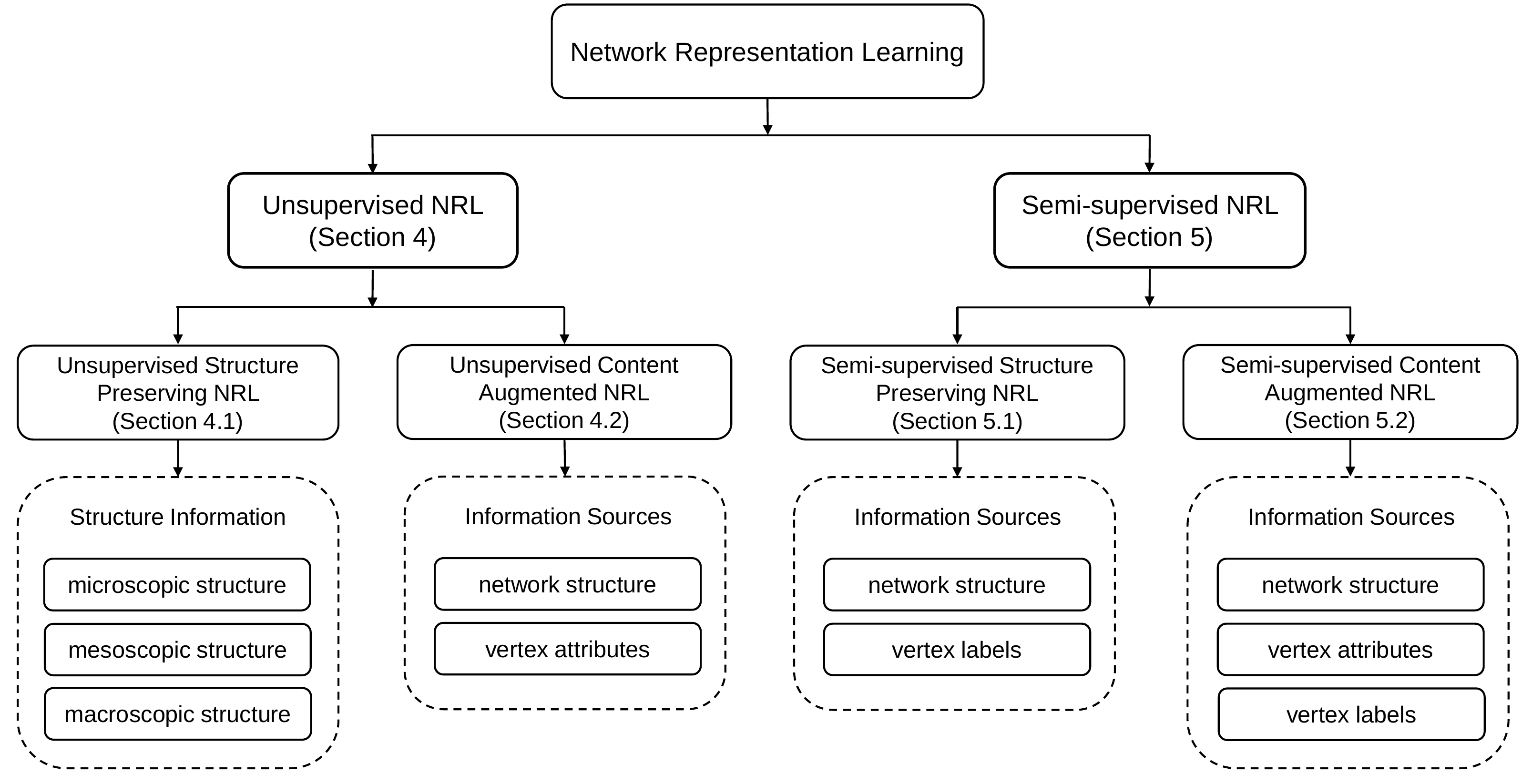}}
	\caption{The proposed taxonomy to summarize network representation learning techniques. We categorize network representation learning into two groups, \textit{unsupervised network representation learning} and \textit{semi-supervised network representation learning}, depending on whether vertex labels are available for learning. For each group, we further categorize methods into two subgroups, depending on whether the representation learning is based on network topology structure only, or augmented with information from node content.}
	\label{fig:taxonomy}
\end{figure*}

Fig.~\ref{fig:NRL} demonstrates a conceptual view of network representation learning, using a toy network. In this case, only network structure is considered to learn vertex representations. Given an information network shown in Fig.~\ref{fig:NRL:subfig:network}, the objective of NRL is to embed all network vertices into a low-dimensional space, as depicted in Fig.~\ref{fig:NRL:subfig:embedding}. In the embedding space, vertices with structural proximity are represented closely to each other. For example, as vertex 7 and vertex 8 are directly connected, the first-order proximity enforces them close to each other in the embedding space. Though vertex 2 and vertex 5 are not directly connected, they are also embedded closely to each other because they have high second-order proximity, which is reflected by 4 common neighbors shared by these two vertices. Vertex 20 and vertex 25 are not directly connected, nor do they share common direct neighbors. However, they are connected by many $k$-step paths $(k\geq 3)$, which proves that they have high-order proximity. Thus, vertex 20 and vertex 25 also have close embeddings. Different from other vertices, vertex 10--16 clearly belong to the same community in the original network. This intra-community proximity guarantees the images of these vertices also exhibit a clear cluster structure in the embedding space.

\section{Categorization}
\label{sec:taxonomy}

In this section, we propose a new taxonomy to categorize existing network representation learning techniques in the literature, as shown in Fig.~\ref{fig:taxonomy}. The first layer of the taxonomy is based on whether vertex labels are provided for learning. According to this, we categorize network representation learning into two groups: \textit{unsupervised network representation learning} and \textit{semi-supervised network representation learning}.

\begin{scriptsize}
	\begin{table*}[tb]
		\centering
		\scriptsize
		\tabcolsep 3pt
		\renewcommand\arraystretch{1.4}
		\caption{A summary of NRL algorithms according to the information sources they use for learning}
		\begin{tabular}{|c|c|c|c|c|c|c|c|}
			\hline
			\multirow{4}{*}{\bf Category} &\multirow{4}{*}{\bf Algorithms} &\multicolumn{4}{c|}{\bf Network Structure} & \multirow{4}{*}{\bf Vertex Attributes} & \multirow{4}{*}{\bf Vertex Labels}\\
			\cline{3-6}
			&&\multirow{3}{*}{Microscopic} & \multicolumn{2}{c|}{Mesoscopic}  & \multirow{3}{*}{Macroscopic} &&\\\cline{4-5}
			& & & Structural Role  & Intra-community  & & &\\
			& & & Proximity  & Proximity  & & &\\
			\hline
			\multirow{19}{*}{\bf Unsupervised}&Social Dim. \cite{tang2009relational,tang2011leveraging,tang2009scalable} & & &\checkmark & & & \\\cline{2-8}
			&DeepWalk \cite{perozzi2014deepwalk} &\checkmark & & & & & \\\cline{2-8}
			&LINE \cite{tang2015line} &\checkmark & & & & & \\\cline{2-8}
			&GraRep \cite{cao2015grarep} &\checkmark & & & & & \\\cline{2-8}
			&DNGR \cite{cao2016deep} &\checkmark & & & & & \\\cline{2-8}
			&SDNE \cite{wang2016structural} &\checkmark & & & & & \\\cline{2-8}
			&node2vec \cite{grover2016node2vec} &\checkmark & & & & & \\\cline{2-8}
			&HOPE \cite{ou2016asymmetric} &\checkmark & & & & & \\\cline{2-8}
			&APP \cite{zhou2017scalable} &\checkmark & & & & & \\\cline{2-8}
			&M-NMF \cite{wang2017community} &\checkmark & &\checkmark & & & \\\cline{2-8}
			&GraphGAN \cite{wang2018graphgan} &\checkmark & & & & &\\\cline{2-8}
			&struct2vec \cite{ribeiro2017struc2vec} & &\checkmark & & & &\\\cline{2-8}
			&GraphWave \cite{donnat2017spectral} & &\checkmark  && & &\\\cline{2-8}
			&SNS \cite{lyu2017enhancing}&\checkmark &\checkmark & & & &\\\cline{2-8}
			&DP \cite{feng2018representation} &\checkmark & & &\checkmark & & \\\cline{2-8}
			&HARP \cite{chen2018harp} &\checkmark & & &\checkmark & & \\\cline{2-8}
			&TADW \cite{yang2015network} &\checkmark & & & &\checkmark & \\\cline{2-8}
			&HSCA \cite{zhang2016collective} &\checkmark & & & &\checkmark & \\\cline{2-8}
			&pRBM \cite{wang2016paired} &\checkmark & & & &\checkmark & \\\cline{2-8}
			&UPP-SNE \cite{zhang2017user} &\checkmark& & & &\checkmark &\\\cline{2-8}
			&PPNE \cite{li2017ppne} &\checkmark & & & &\checkmark &\\\hline
			\multirow{10}{*}{\bf Semi-supervised}&DDRW \cite{li2016discriminative} &\checkmark & & & & &\checkmark \\\cline{2-8}
			&MMDW \cite{tu2016max} &\checkmark & & & & &\checkmark \\\cline{2-8}
			&TLINE \cite{zhang2016tline} &\checkmark & & & & &\checkmark \\\cline{2-8}
			&GENE \cite{chen2016incorporate} &\checkmark & & & & &\checkmark \\\cline{2-8}
			&SemiNE \cite{li2017semi} &\checkmark & & & & &\checkmark\\\cline{2-8}
			&TriDNR \cite{pan2016tri} &\checkmark & & & &\checkmark &\checkmark \\\cline{2-8}
			&LDE \cite{wang2016linked}  &\checkmark & & & &\checkmark &\checkmark \\\cline{2-8}
			&DMF \cite{zhang2016collective} &\checkmark & & & &\checkmark &\checkmark \\\cline{2-8}
			&Planetoid \cite{yang2016revisiting} &\checkmark & & & &\checkmark &\checkmark \\\cline{2-8}
			&LANE \cite{huang2017label} &\checkmark & & & &\checkmark &\checkmark \\\hline
		\end{tabular}
		\label{table-infosource}
	\end{table*}
	%\end{sidewaystable}
\end{scriptsize}

\noindent\textbf{Unsupervised network representation learning}. In this setting, there are no labeled vertices provided for learning vertex representations. Network representation learning is therefore considered as a generic task independent of subsequent learning, and vertex representations are learned in an unsupervised manner.
	
Most of the existing NRL algorithms fall into this category. After vertex representations are learned in a new embedding space, they are taken as features to any vector-based algorithms for various learning tasks. Unsupervised NRL algorithms can be further divided into two subgroups based on the type of network information available for learning: unsupervised structure preserving methods that preserve only network structure, and unsupervised content augmented methods that incorporate vertex attributes and network structure to learn joint vertex embeddings.
	
\noindent\textbf{Semi-supervised network representation learning}. In this case, there exist some labeled vertices for representation learning. Because vertex labels play an essential role in determining the categorization of each vertex with strong correlations to network structure and vertex attributes, semi-supervised network representation learning is proposed to take advantage of vertex labels available in the network for seeking more effective joint vector representations.
	
In this setting, network representation learning is coupled with supervised learning tasks such as vertex classification. A unified objective function is often formulated to simultaneously optimize the learning of vertex representations and the classification of network vertices. Therefore, the learned vertex representations can be both informative and discriminative with respect to different categories. Semi-supervised NRL algorithms can also be categorized into two subgroups, semi-supervised structure preserving methods and semi-supervised content augmented methods.

Table~\ref{table-infosource} summarizes all NRL algorithms, according to the information sources that they use for representation learning. In general, there are three main types of information sources: network structure, vertex attributes, and vertex labels. Most of the unsupervised NRL algorithms focus on preserving network structure for learning vertex representations, and only a few algorithms (\textit{e.g.}, TADW~\cite{yang2015network}, HSCA~\cite{zhang2016collective}) attempt to leverage vertex attributes. 
By contrast, under the semi-supervised learning setting, half of the algorithms intend to couple vertex attributes with network structure and vertex labels to learn vertex representations. On both settings, most of the algorithms focus on preserving microscopic structure, while very few algorithms (\textit{e.g.}, M-NMF~\cite{wang2017community}, DP~\cite{feng2018representation}, HARP~\cite{chen2018harp}) attempt to take advantage of the mesoscopic and macroscopic structure. 

\begin{scriptsize}
	\begin{table*}[tb]
		\renewcommand\arraystretch{1.4}
		\centering
		\scriptsize
		\caption{A categorization of NRL algorithms from methodology perspectives}
		\begin{tabular}{|c|c|c|c|}
			\hline
			\bf Methodology & \bf Algorithms & \bf Advantage & \bf Disadvantage \\\hline
			\bf Matrix Factorization & 
			\tabincell{c}{
				Social Dim. \cite{tang2009relational,tang2011leveraging},
				GraRep \cite{cao2015grarep},
				HOPE \cite{ou2016asymmetric},\\
				GraphWave\cite{donnat2017spectral},
				M-NMF \cite{wang2017community},
				TADW \cite{yang2015network},
				HSCA \cite{zhang2016homophily},\\
				MMDW \cite{tu2016max},
				DMF \cite{zhang2016collective},
				LANE \cite{huang2017label}} & capture global structure & high time and memory cost \\\hline
			\bf Random Walk & 
			\tabincell{c}{
				DeepWalk \cite{perozzi2014deepwalk},
				node2vec \cite{grover2016node2vec},
				APP \cite{zhou2017scalable},
				DDRW \cite{li2016discriminative},\\
				GENE \cite{chen2016incorporate},
				TriDNR \cite{pan2016tri},
				UPP-SNE \cite{zhang2017user},
				struct2vec \cite{ribeiro2017struc2vec},\\
				SNS \cite{lyu2017enhancing},
				PPNE \cite{li2017ppne},
				SemiNE \cite{li2017semi}}
				 & relatively efficient & only capture local structure \\\hline
			\bf Edge Modeling & 
			\tabincell{c}{
				LINE \cite{tang2015line},
				TLINE \cite{zhang2016tline},
				LDE \cite{wang2016linked},
				pRBM \cite{wang2016paired},\\
				GraphGAN\cite{wang2018graphgan}} & efficient & only capture local structure \\\hline
			\bf Deep Learning & DNGR \cite{cao2016deep},
			SDNE \cite{wang2016structural} & capture non-linearity & high time cost \\\hline
			\bf Hybrid & DP \cite{feng2018representation},
			HARP \cite{chen2018harp}, Planetoid \cite{yang2016revisiting} & capture global structure &\\\hline
		\end{tabular}
		\label{table-methodology}
	\end{table*}
\end{scriptsize}

Approaches to network representation learning in the above two different settings can be summarized into five categories from algorithmic perspectives.
\begin{enumerate}
	\item \textbf{Matrix factorization based methods}. Matrix factorization based methods represent the connections between network vertices in the form of a matrix and use matrix factorization to obtain the embeddings. Different types of matrices are constructed to preserve network structure, such as the $k$-step transition probability matrix, the modularity matrix, or the vertex-context matrix \cite{yang2015network}. By assuming that such high-dimensional vertex representations are only affected by a small quantity of latent factors, matrix factorization is used to embed the high-dimensional vertex representations into a latent, low-dimensional structure preserving space.
	
	Factorization strategies vary across different algorithms according to their objectives. For example, in the Modularity Maximization method \cite{tang2009relational}, eigen decomposition is performed on the modularity matrix to learn community indicative vertex representations~\cite{newman2006finding}; in the TADW algorithm~\cite{yang2015network}, inductive matrix factorization \cite{natarajan2014inductive} is carried out on the vertex-context matrix to simultaneously preserve vertex textual features and network structure in the learning of vertex representations. Although matrix factorization based methods have been proved effective in learning informative vertex representations, the scalability is a major bottleneck because carrying out factorization on a matrix with millions of rows and columns is memory intensive and computationally expensive or, sometime, even infeasible.
	
	\item \textbf{Random walk based methods}. For scalable vertex representation learning, random walk is exploited to capture structural relationships between vertices. By performing truncated random walks, an information network is transformed into a collection of vertex sequences, in which, the occurrence frequency of a vertex-context pair measures the structural distance between them. Borrowing the idea of word representation learning~\cite{mikolov2013efficient,mikolov2013distributed}, vertex representations are then learned by using each vertex to predict its contexts. DeepWalk \cite{perozzi2014deepwalk} is the pioneer work in using random walks to learn vertex representations. node2vec~\cite{grover2016node2vec} further exploits a biased random walk strategy to capture more flexible contextual structure. 
	
	As the extensions of the structure only preserving version, algorithms like DDRW~\cite{li2016discriminative}, GENE~\cite{chen2016incorporate} and SemiNE~\cite{li2017semi} incorporate vertex labels with network structure to harness representation learning, PPNE~\cite{li2017ppne} imports vertex attributes, and Tri-DNR \cite{pan2016tri} enforces the model with both vertex labels and attributes. As these models can be trained in an on-line manner, they have great potential to scale up. 
	
	\item \textbf{Edge modeling based methods}. Different from approaches that use matrix or random walk to capture network structure, the edge modeling based methods directly learn vertex representations from vertex-vertex connections. For capturing the first-order and second-order proximity, LINE~\cite{tang2015line} models a joint probability distribution and a conditional probability distribution, respectively, on connected vertices. To learn the representations of linked documents, LDE~\cite{wang2016linked} models the document-document relationships by maximizing the conditional probability between connected documents. pRBM~\cite{wang2016paired} adapts the RBM~\cite{hinton2006reducing} model to linked data by making the hidden RBM representations of connected vertices similar to each other. GraphGAN~\cite{wang2018graphgan} adopts Generative Adversarial Nets (GAN)~\cite{goodfellow2014generative} to accurately model the vertex connectivity probability. Edge modeling based methods are more efficient compared to matrix factorization and random walk based methods. However, these methods cannot capture global network structure as they only consider observable vertex connectivity information. 
	
	\item \textbf{Deep learning based methods}. To extract complex structure features and learn deep, highly non-linear vertex representations, deep learning techniques~\cite{salakhutdinov2009semantic,vincent2010stacked} are also applied to network representation learning. For example, DNGR~\cite{cao2016deep} applies the \textit{stacked denoising autoencoders} (SDAE)~\cite{vincent2010stacked} on the high-dimensional matrix representations to learn deep low-dimensional vertex representations. SDNE~\cite{wang2016structural} uses a semi-supervised deep autoencoder model~\cite{salakhutdinov2009semantic} to model non-linearity in network structure. Deep learning based methods have the ability to capture non-linearity in networks, but their computational time cost is usually high. Traditional deep learning architectures are designed for 1D, 2D, or 3D Euclidean structured data, but efficient solutions need to be developed on non-Euclidean structured data like graphs.
	
	%their scalability and interpretability need to be further investigated.

%SDNE~\cite{wang2016structural} is realized by a semi-supervised deep autoencoder model \cite{salakhutdinov2009semantic}, in which the unsupervised component reconstructs the second-order proximity to retain the global network structure, while the supervised component exploits the first-order proximity as supervised information to preserve the local network structure. 

	\item \textbf{Hybrid methods.} Some other methods make use of a mixture of above methods to learn vertex representations. For example, DP~\cite{feng2018representation} enhances spectral embedding~\cite{belkin2002laplacian} and DeepWalk~\cite{perozzi2014deepwalk} with the degree penalty principle to preserve the macroscopic scale-free property. HARP~\cite{chen2018harp} takes advantage of random walk based methods (DeepWalk~\cite{perozzi2014deepwalk} and node2vec~\cite{grover2016node2vec}) and edge modeling based method (LINE~\cite{tang2015line}) to learn vertex representations from small sampled networks to the original network.
\end{enumerate}

We summarize all five categories of network representation learning techniques and compare their advantages and disadvantages in Table~\ref{table-methodology}.

\section{Unsupervised Network Representation Learning}
\label{sec:unsupervised}

In this section, we review unsupervised network representation learning methods by separating them into two subsections, as outlined in Fig.~\ref{fig:taxonomy}. After that, we summarize key characteristics of the methods and compare their differences across the two categories.

\subsection{Unsupervised Structure Preserving Network Representation Learning}
\label{sec:unsupervised-structure}

Structure preserving network representation learning refers to methods that intend to preserve network structure, in the sense that vertices close to each other in the original network space should be represented similarly in the new embedding space. In this category, research efforts have been focused on designing various models to capture structure information conveyed by the original network as much as possible. 

\begin{figure}[!htbp]
	\centering
	\scalebox{0.3}{
		\includegraphics[width=\textwidth]{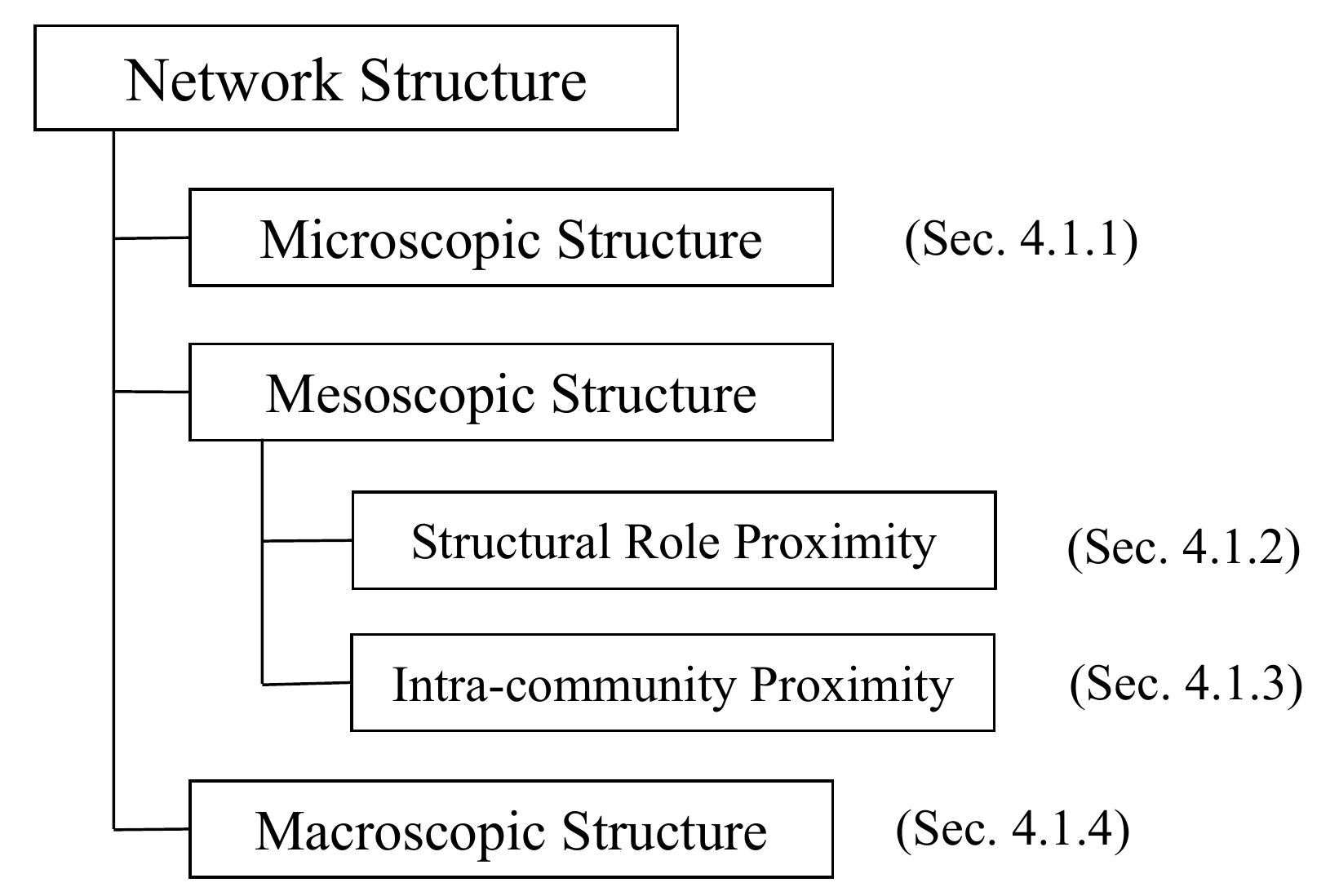}}
	\caption{Categorization of network structure.}
	\label{fig:4.1struct}
\end{figure}

We summarize network structure considered for learning vertex representations into three types: (i) \textbf{microscopic structure}, which includes local closeness proximity, i.e., the first-order, second-order, and high-order proximity, (ii) \textbf{mesoscopic structure}, which captures structural role proximity and the intra-community proximity, and (iii) \textbf{macroscopic structure}, which captures global network properties, such as the scale-free property or small world property. The following subsections are organized according to our categorization of network structure, as depicted in Fig.~\ref{fig:4.1struct}.

\subsubsection{Microscopic Structure Preserving NRL}

This category of NRL algorithms aim to preserve local structure information among directly or indirectly connected vertices in their neighborhood, including first-order, second-order, and high-order proximity. The first-order proximity captures the homophily, i.e., directly connected vertices tend to be similar to each other, while the second-order and high-order proximity captures the similarity between vertices sharing common neighbors. Most of structure preserving NRL algorithms fall into this category.

\begin{figure}[tb]
	\centering
	\scalebox{0.4}{
		\includegraphics[width=\textwidth]{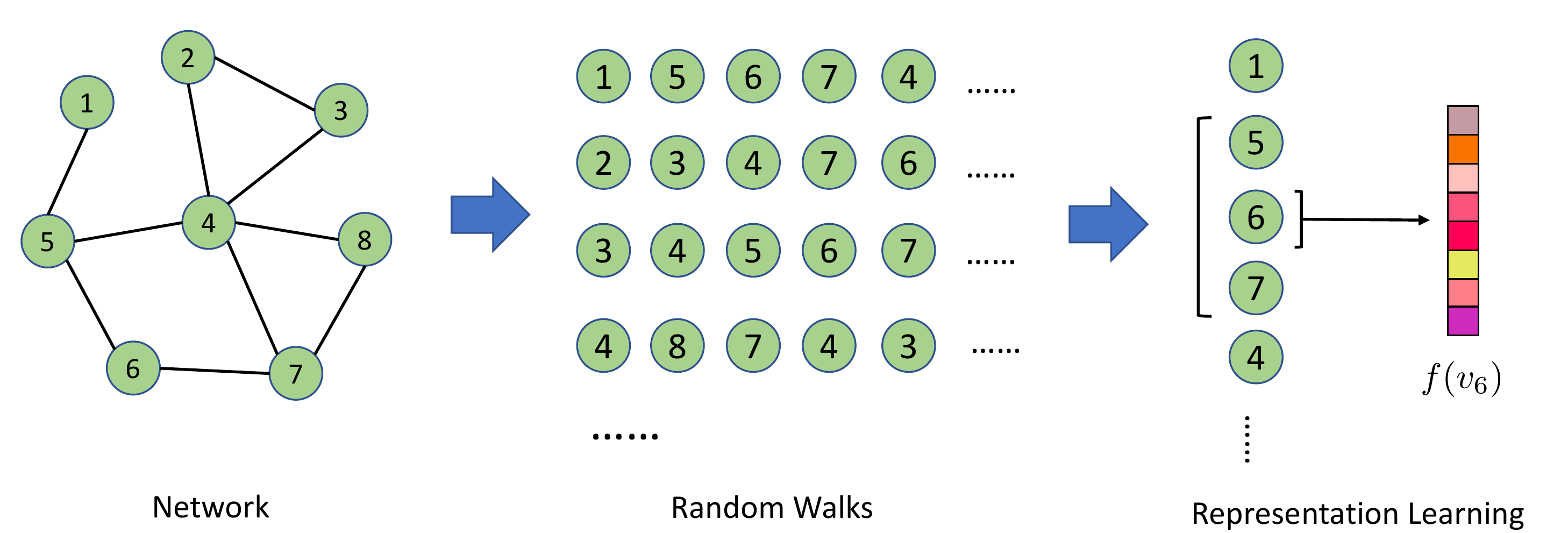}}
	\caption{The workflow of DeepWalk. It first generates random walk sequences from a given network, and then applies the Skip-Gram model to learn vertex representations.}
	\label{fig:deepwalk}
\end{figure}

\textbf{DeepWalk}. DeepWalk~\cite{perozzi2014deepwalk} generalizes the idea of the Skip-Gram model~\cite{mikolov2013efficient,mikolov2013distributed} that utilizes word context in sentences to learn latent representations of words, to the learning of latent vertex representations in networks, by making an analogy between natural language sentence and short random walk sequence. The workflow of DeepWalk is given in
Fig. \ref{fig:deepwalk}. Given a random walk sequence with length $L$, $\lbrace v_{1}, v_{2}, \cdots, v_{L}\rbrace$,
following Skip-Gram, DeepWalk learns the representation of vertex $v_{i}$ by using it to predict its context vertices, which is achieved by the optimization problem:
\begin{equation}
\small
\min_{f}\;-\log \mathrm{Pr}(\left\lbrace v_{i-t},\cdots, v_{i+t}\right\rbrace\setminus v_{i}\vert f(v_{i})),
\end{equation}
where $\left\lbrace v_{i-t},\cdots, v_{i+t}\right\rbrace\setminus v_{i}$ are the context vertices of vertex $v_{i}$ within $t$ window size. Making conditional independence assumption, the probability $\mathrm{Pr}(\left\lbrace v_{i-t},\cdots, v_{i+t}\right\rbrace\setminus v_{i}\vert f(v_{i}))$ is approximated as
\begin{equation}
\small
\mathrm{Pr}\left( \left\lbrace v_{i-t},\cdots, v_{i+t}\right\rbrace\setminus v_{i}\vert f(v_{i})\right)=\prod_{j=i-t,j\neq i}^{i+t}\mathrm{Pr}(v_{j}\vert f(v_{i})).
\end{equation}
%To speed up the training, Hierarchical Softmax \cite{mnih2009scalable,morin2005hierarchical} is used to model the probability $\mathrm{Pr}(v_{j}\vert f(v_{i}))$.

Following the DeepWalk's learning architecture, vertices that share similar context vertices in random walk sequences should  be represented closely in the new embedding space. Considering the fact that context vertices in random walk sequences describe neighborhood structure, DeepWalk actually represents vertices sharing similar neighbors (direct or indirect) closely in the embedding space, so the second-order and high-order proximity is preserved. 

\textbf{Large-scale Information Network Embedding (LINE)}. Instead of exploiting random walks to capture network structure, LINE~\cite{tang2015line} learns vertex representations by explicitly modeling the first-order and second-order proximity. To preserve the first-order proximity, LINE minimizes the following objective:
\begin{equation} \label{LINE_obj1}\small
O_{1}=d(\hat{p}_{1}(\cdot,\cdot),p_{1}(\cdot,\cdot)).
\end{equation}
For each vertex pair $v_{i}$ and $v_{j}$ with $(v_{i},v_{j})\in E$, $p_{1}(\cdot,\cdot)$ is the joint distribution modeled by their latent embeddings $r_{v_{i}}$ and $r_{v_{j}}$. $\hat{p}_{1}(v_{i},v_{j})$ is the empirical distribution between them. $d(\cdot,\cdot)$ is the distance between two distributions. 

To preserve the second-order proximity, LINE minimizes the following objective:
\begin{equation} \label{LINE_obj2}\small
O_{2}=\sum_{v_{i}\in V}{\lambda}_{i}d(\hat{p}_{2}(\cdot|v_{i}),p_{2}(\cdot|v_{i})),
\end{equation}where $p_{2}(\cdot|v_{i})$ is the context conditional distribution for each $v_{i}\in V$ modeled by vertex embeddings, $\hat{p}_{2}(\cdot|v_{i})$ is the empirical conditional distribution and ${\lambda}_{i}$ is the prestige of vertex $v_{i}$. Here, vertex context is determined by its neighbors, i.e., for each $v_{j}$, $v_{j}$ is $v_{i}$'s context, if and only if $(v_{i},v_{j})\in E$.

By minimizing these two objectives, LINE learns two kinds of vertex representations that preserve the first-order and second-order proximity, and takes their concatenation as the final vertex representation. 

\textbf{GraRep}. Following the idea of DeepWalk~\cite{perozzi2014deepwalk}, GraRep~\cite{cao2015grarep} extends the skip-gram model to capture the high-order proximity, i.e., vertices sharing common $k$-step neighbors $(k \geq 1)$ should have similar latent representations. Specifically, for each vertex, GraRep defines its $k$-step neighbors $(k \geq 1)$ as context vertices, and for each $1\leq k\leq K$, to learn $k$-step vertex representations, GraRep employs the matrix factorization version of skip-gram:
\begin{equation}\small
\left[U^{k},\Sigma^{k},V^{k}\right]=SVD(X^{k}).
\end{equation}where $X^{k}$ is the log $k$-step transition probability matrix. The $k$-step representation for vertex $v_{i}$ is constructed as the $i$th row of matrix $U^{k}_{d}(\Sigma^{k}_{d})^{\frac{1}{2}}$, where $U^{k}_{d}$ is the first-$d$ columns of $U^{k}$ and $\Sigma^{k}_{d}$ is the diagonal matrix composed of the top $d$ singular values. After $k$-step vertex representations are learned, GraRep concatenates them together as the final vertex representations.

\textbf{Deep Neural Networks for Graph Representations (DNGR)}. To overcome the weakness of truncated random walks in exploiting vertex contextual information, i.e., the difficulty in capturing correct contextual information for vertices at the boundary of sequences and the difficulty in determining the walk length and the number of walks, DNGR~\cite{cao2016deep} utilizes the random surfing model to capture contextual relatedness between each pair of vertices and preserves them into $|V|$-dimensional vertex representations $X$. To extract complex features and model non-linearities, DNGR applies the \textit{stacked denoising autoencoders} (SDAE) \cite{vincent2010stacked} to the high-dimensional vertex representations $X$ to learn deep low-dimensional vertex representations.

\textbf{Structural Deep Network Embedding (SDNE)}. SDNE~\cite{wang2016structural} is a deep learning based approach that uses a semi-supervised deep autoencoder model to capture non-linearity in network structure. In the unsupervised component, SDNE learns the second-order proximity preserving vertex representations via reconstructing the $|V|$-dimensional vertex adjacent matrix representations, which tries to minimize
\begin{equation} \label{SDNE_obj_2nd}\small
\mathcal{L}_{2nd}=\sum_{i=1}^{|V|}\|(r^{(0)}_{v_{i}}-\hat{r}^{(0)}_{v_{i}})\odot\bm{b}_{i}\|_{2}^{2},
\end{equation}where $r^{(0)}_{v_{i}}=S_{i:}$ is the input representation and $\hat{r}^{(0)}_{v_{i}}$ is the reconstructed representation. $\bm{b}_{i}$ is a weight vector used to penalize construction error more on non-zero elements of $S$. 

%In the above objective, $\odot$ means the Hadamard product, $\bm{b}_{i}=\{b_{ij}\}^{|V|}_{j=1}$, with $b_{ij}=1$ for $S_{ij}=0$ and $b_{ij}=\beta>1$ for $S_{ij}\neq 0$.

In the supervised component, SDNE imports the first-order proximity by penalizing the distance between connected vertices in the embedding space. The loss function for this objective is defined as:
\begin{equation} \label{SDEN_obj_1st}\small
\mathcal{L}_{1st}=\sum_{i,j=1}^{|V|}S_{ij}\|r^{(K)}_{v_{i}}-r^{(K)}_{v_{j}}\|_{2}^{2},
\end{equation}where $r^{(K)}_{v_{i}}$ is the $K$-th layer representation of vertex $v_{i}$, with $K$ being the number of hidden layers.

In all, SDNE minimizes the joint objective function:
\begin{equation} \label{SDNE_overall_obj}
\mathcal{L}=\mathcal{L}_{2nd}+\alpha\mathcal{L}_{1st}+\nu\mathcal{L}_{reg},
\end{equation}where $\mathcal{L}_{reg}$ is a regularization term to prevent overfitting. %, which is defined as
%\begin{equation}
%\mathcal{L}_{reg}=\frac{1}{2}\sum_{k=1}^{K}(\|W^{(k)}\|^{2}_{F}+\|\hat{W}^{(k)}\|^{2}_{F}).
%\end{equation}
After solving the minimization of (\ref{SDNE_overall_obj}), for vertex $v_{i}$, the $K$-th layer representation $r^{(K)}_{v_{i}}$ is taken as its representation $r_{v_{i}}$.

\begin{figure}[tb]
	\centering
	\scalebox{0.32}{
		\includegraphics[width=\textwidth]{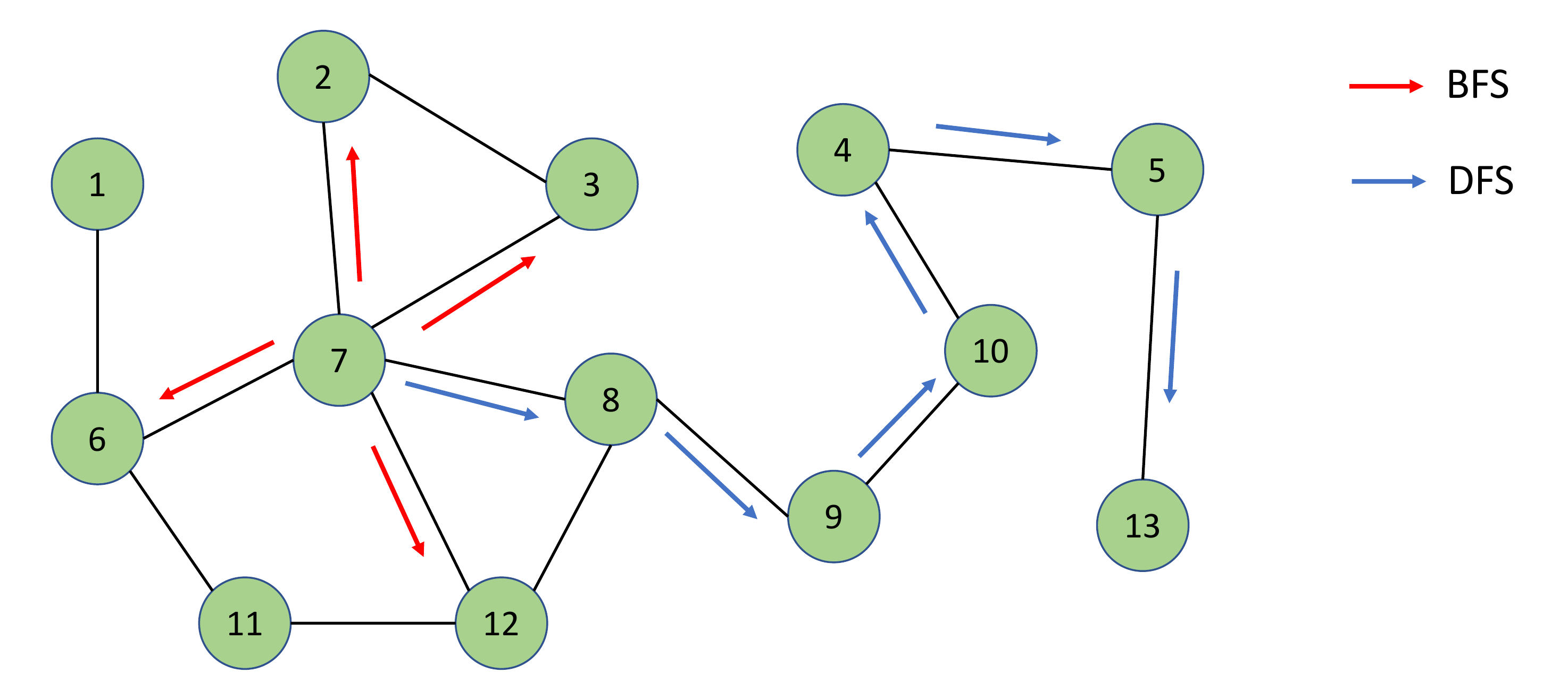}}
	\caption{Two different neighborhood sampling strategies considered by node2vec: BFS and DFS.}
	\label{fig:node2vec}
\end{figure}

\textbf{node2vec}. In contrast to the rigid strategy of defining neighborhood (context) for each vertex, node2vec~\cite{grover2016node2vec} designs a flexible neighborhood sampling strategy, i.e., biased random walk, which smoothly interpolates between two extreme sampling strategies, i.e., Breadth-first Sampling (BFS) and Depth-first Sampling (DFS), as illustrated in Fig. \ref{fig:node2vec}. The biased random walk exploited in node2vec can better preserve both the second-order and high-order proximity. 

Following the skip-gram architecture, given the set of neighbor vertices $N(v_{i})$ generated by biased random walk, node2vec learns the vertex representation $f(v_{i})$ by optimizing the occurrence probability of neighbor vertices $N(v_{i})$ conditioned on the representation of vertex $v_{i}$, $f(v_{i})$:
\begin{equation} \label{node2vec_obj}\small
\max_{f}\sum_{v_{i}\in V}\log\mathrm{Pr}(N(v_{i})|f(v_{i})).
\end{equation}

\textbf{High-order Proximity Preserved Embedding (HOPE)}. HOPE~\cite{ou2016asymmetric} learns vertex representations that capture the asymmetric high-order proximity in directed networks. In undirected networks, the transitivity is symmetric, but it is asymmetric in directed networks. For example, in an directed network, if there is a directed link from vertex $v_{i}$ to vertex $v_{j}$ and from vertex $v_{j}$ to vertex $v_{k}$, it is more likely to have a directed link from $v_{i}$ to $v_{k}$, but not from $v_{k}$ to $v_{i}$.

To preserve the asymmetric transitivity, HOPE learns two vertex embedding vectors $U^{s},U^{t}\in\mathbb{R}^{|V|\times d}$, which is called source and target embedding vectors, respectively. 
After constructing the high-order proximity matrix $S$ from four proximity measures, i.e., Katz Index \cite{katz1953new}, Rooted PageRank \cite{song2009scalable}, Common Neighbors and Adamic-Adar. HOPE learns vertex embeddings by solving the following matrix factorization problem:
\begin{equation}\label{HOPE_obj}\small
\min_{U_{s},U_{t}}\Vert S-U^{s}\cdot{U^{t}}^{\mathrm{T}}\Vert_{F}^{2}.
\end{equation}

\textbf{Asymmetric Proximity Preserving graph embedding (APP)}. APP~\cite{zhou2017scalable} is another NRL algorithm designed to capture asymmetric proximity, by using a Monte Carlo approach to approximate the asymmetric Rooted PageRank proximity~\cite{song2009scalable}. Similar to HOPE, APP has two representations for each vertex $v_{i}$, the one as a source role $r^{s}_{v_{i}}$ and the other as a target role $r^{t}_{v_{i}}$. For each sampled path starting from $v_{i}$ and ending with $v_{j}$, the representations are learned by maximizing the target vertex $v_{j}$'s occurrence probability conditioned on the source vertex $v_{i}$:
\begin{equation}\small
\mathrm{Pr}(v_{j}|v_{i})=\frac{\exp(r^{s}_{v_{i}}\cdot r^{t}_{v_{j}})}{\sum_{v\in V}\exp(r^{s}_{v_{i}}\cdot r^{t}_{v})}.
\end{equation}

\textbf{GraphGAN}. GraphGAN~\cite{wang2018graphgan} learns vertex representations by modeling the connectivity behavior through an adversarial learning framework. Inspired by GAN (Generative Adversarial Nets)~\cite{goodfellow2014generative}, GraphGAN works through two components: (i) Generator $G(v|v_{c})$, which fits the distribution of the vertices connected to $v_{c}$ across $V$ and generates the likely connected vertices, and (ii) Discriminator $D(v,v_{c})$, which outputs a connecting probability for the vertex pair $(v,v_{c})$, to differentiate the vertex pairs generated by $G(v|v_{c})$ from the ground truth. $G(v|v_{c})$ and $D(v,v_{c})$ compete in a way that $G(v|v_{c})$ tries to fit the true connecting distribution as much as possible and generates fake connected vertex pairs to fool $D(v,v_{c})$, while $D(v,v_{c})$ tries to increase its discriminative power to distinguish the vertex pairs generated by $G(v|v_{c})$ from the ground truth. The competition is achieved by the following $\textit{minimax}$ game:
\begin{equation}\label{GraphGAN_obj}\small
\begin{aligned}
\min_{\theta_{G}}\max_{\theta_{D}}\sum_{v_{c}\in V}&\left(\mathbb{E}_{v\sim\mathrm{Pr}_{true}(\cdot|v_{c})}\left[\log D(v,v_{c};\theta_{D})\right]\right.\\
&+\left.\mathbb{E}_{v\sim G(\cdot|v_{c};\theta_{G})}\left[\log(1-D(v,v_{c};\theta_{D}))\right]\right).
\end{aligned}
\end{equation}Here, $G(v|v_{c};\theta_{G})$ and $D(v,v_{c};\theta_{D})$ are defined as following:
\begin{equation}\small
\begin{aligned}
G(v|v_{c};\theta_{G})=\frac{\exp(\mathbf{g}_{v}\cdot\mathbf{g}_{v_{c}})}{\sum_{v\neq v_{c}}\exp(\mathbf{g}_{v}\cdot\mathbf{g}_{v_{c}})},\\
D(v,v_{c};\theta_{D})=\frac{1}{1+\exp(\mathbf{d}_{v}\cdot\mathbf{d}_{v_{c}})},
\end{aligned}
\end{equation}where $\mathbf{g}_{v}\in\mathbb{R}^{k}$ and $\mathbf{d}_{v}\in\mathbb{R}^{k}$ is the representation vector for generator and discriminator, respectively, and $\theta_{D}=\{\mathbf{d}_{v}\}$, $\theta_{G}=\{\mathbf{g}_{v}\}$. After the $\textit{minimax}$ game in Eq. (\ref{GraphGAN_obj}) is solved, $\mathbf{g}_{v}$ serves as the final vertex representations.

\begin{scriptsize}
	\begin{table}[tb]
		\centering
		\scriptsize
		\tabcolsep 3pt
		\renewcommand\arraystretch{1.3}
		\caption{A summary of microscopic structure preserving NRL algorithms}
		\begin{tabular}{|c|c|c|c|}
			\hline
			\multirow{2}{*}{\bf Algorithms} &First-order   & Second-order & High-order \\
			& Proximity & Proximity & Proximity \\\hline
			DeepWalk \cite{perozzi2014deepwalk} & &\checkmark &\checkmark \\\hline
		    LINE \cite{tang2015line} &\checkmark &\checkmark & \\\hline
			GraRep \cite{cao2015grarep} & &\checkmark &\checkmark \\\hline
			DNGR \cite{cao2016deep} & &\checkmark &\checkmark \\\hline
			SDNE \cite{wang2016structural} &\checkmark &\checkmark &  \\\hline
			node2vec \cite{grover2016node2vec} & &\checkmark &\checkmark \\\hline
			HOPE \cite{ou2016asymmetric} & &\checkmark &\checkmark \\\hline
			APP \cite{zhou2017scalable} & &\checkmark &\checkmark \\\hline
			GraphGAN \cite{wang2018graphgan} &\checkmark & & \\\hline
		\end{tabular}
		\label{table-localcloseness}
	\end{table}
\end{scriptsize}

\textbf{Summary}: The proximity preserved by microscopic structure preserving NRL algorithms is summarized in Table \ref{table-localcloseness}. Most algorithms in this category preserve the second-order and high-order proximity, whereas only LINE \cite{tang2015line}, SDNE \cite{wang2016structural} and GraphGAN \cite{wang2018graphgan} consider the first-order proximity. From the methodology perspective, DeepWalk \cite{perozzi2014deepwalk}, node2vec \cite{grover2016node2vec} and APP \cite{zhou2017scalable} employ random walks to capture vertex neighborhood structure. GraRep \cite{cao2015grarep} and HOPE \cite{ou2016asymmetric} are realized by performing factorization on a $|V|\times|V|$ scale matrix, making them hard to scale up. LINE~\cite{tang2015line} and GraphGAN~\cite{wang2018graphgan} directly model the connectivity behavior, while deep learning based methods (DNGR~\cite{cao2016deep} and SDNE~\cite{wang2016structural}) learn non-linear vertex representations. 

\subsubsection{Structural Role Proximity Preserving NRL}
	
Besides local connectivity patterns, vertices often share similar structural roles at a mesoscopic level, such as centers of stars or members of cliques. Structural role proximity preserving NRL aims to embed vertices that are far away from each other but share similar structural roles close to each other. This not only facilitates the downstream structural role dependent tasks but also enhances microscopic structure preserving NRL.
	
\textbf{struct2vec}. struct2vec~\cite{ribeiro2017struc2vec} first encodes the vertex structural role similarity into a multilayer graph, where the weights of edges at each layer are determined by the structural role difference at the corresponding scale. DeepWalk~\cite{perozzi2014deepwalk} is then performed on the multilayer graph to learn vertex representations, such that vertices close to each other in the multilayer graph (with high structural role similarity) are embedded closely in the new representation space.
	
For each vertex pair $(v_{i},v_{j})$, considering their $k$-hop neighborhood formed by their neighbors within $k$ steps, their structural distance at scale $k$, $D_{k}(v_{i},v_{j})$, is defined as
\begin{equation}\small
D_{k}(v_{i},v_{j})=D_{k-1}(v_{i},v_{j})+g(s(R_{k}(v_{i})),s(R_{k}(v_{j}))),
\end{equation}where $R_{k}(v_{i})$ is the set of vertices in $v_i$'s $k$-hop neighborhood, $s(R_{k}(v_{i}))$ is the ordered degree sequence of the vertices in $R_{k}(v_{i})$, and $g(s(R_{k}(v_{i})),s(R_{k}(v_{j})))$ is the distance between the ordered degree sequences $s(R_{k}(v_{i}))$ and $s(R_{k}(v_{j}))$. When $k=0$, $D_{0}(v_{i},v_{j})$ is the degree difference between vertex $v_{i}$ and $v_{j}$.
	
\textbf{GraphWave}. By making use of the spectral graph wavelet diffusion patterns, GraphWave~\cite{donnat2017spectral} embeds vertex neighborhood structure into a low-dimensional space and preserves the structural role proximity. The assumption is that, if two vertices residing distantly in the network share similar structural roles, the graph wavelets starting at them will diffuse similarly across their neighbors. 
	
For vertex $v_{k}$, its spectral graph wavelet coefficients $\Psi_{k}$ is defined as
\begin{equation}\small
	\Psi_{k}=U \mathrm{Diag}(g_{s}(\lambda_{1}),\cdots,g_{s}(\lambda_{|V|}))U^{\mathrm{T}} \delta_{k},
\end{equation}where $U$ is the eigenvector matrix of the graph Laplacian $L$ and $\lambda_{1},\cdots,\lambda_{|V|}$ are the eigenvalues, $g_{s}(\lambda)=\exp(-\lambda s)$ is the heat kernel, and $\delta_{k}$ is the one-hot vector for $k$. By taking $\Psi_{k}$ as a probability distribution, the spectral wavelet distribution pattern in $\Psi_{k}$ is then encoded into its empirical characteristic function:
\begin{equation}\small
	\phi_{k}(t)=\frac{1}{|V|}\sum_{m=1}^{|V|}e^{it\Psi_{km}}.
\end{equation}Then $v_{k}$'s low-dimensional representation is then obtained by sampling the 2-dimensional parametric function of $\phi_{k}(t)$ at $d$ evenly separated points $t_{1},t_{2},\cdots,t_{d}$ as:
\begin{equation}
\small
\begin{aligned}
	f(v_{k})=[&\mathrm{Re}(\phi_{k}(t_{1})),\cdots,\mathrm{Re}(\phi_{k}(t_{d})),
	\\&\mathrm{Im}(\phi_{k}(t_{1})),\cdots,\mathrm{Im}(\phi_{k}(t_{d}))].
\end{aligned}
\end{equation}
	
\textbf{Structural and Neighborhood Similarity preserving network embedding (SNS)}. SNS~\cite{lyu2017enhancing} enhances a random walk based method with structural role proximity. To preserve vertex structural roles, SNS represents each vertex as a \textit{Graphlet Degree Vector} with each element being the number of times the given vertex is touched by the corresponding orbit of graphlets. The \textit{Graphlet Degree Vector} is used to measure the vertex structural role similarity. %SNS also utilizes random walks to preserve microscopic structure.
	
Given a vertex $v_{i}$, SNS uses its context vertices $\mathcal{C}(v_{i})$ and structurally similar vertices $\mathcal{S}(v_{i})$ to predict its existence, which is achieved by maximizing the following probability:
\begin{equation}\small
	\mathrm{Pr}(v_{i}|\mathcal{C}(v_{i}),\mathcal{S}(v_{i}))=\frac{\exp(r^{\prime}_{v_{i}}\cdot h_{v_{i}})}{\sum_{u\in V}\exp(r^{\prime}_{u}\cdot h_{v_{i}})},
\end{equation}where $r^{\prime}_{v_{i}}$ is the output representation of $v_{i}$ and $h_{v_{i}}$ is the hidden layer representation for predicting $v_{i}$, which is aggregated from the input representations $r_{u}$, for each $u$ in $\mathcal{C}(v_{i})$ and $\mathcal{S}(v_{i})$. 
	%\mathrm{Pr}(v_{i}|v_{i}\mathrm{'s context vertices}, v_{i}\mathrm{'s structural similar vertices})
	
\textbf{Summary}: struct2vec~\cite{ribeiro2017struc2vec} and GraphWave~\cite{donnat2017spectral} take advantage of structural role proximity to learn vertex representations that facilitate specific structural role dependent tasks, \textit{e.g.}, vertex classification in traffic networks, while SNS~\cite{lyu2017enhancing} enhances a random walk based microscopic structure preserving NRL algorithm with structural role proximity. Technically, random walk is employed by struct2vec and SNS, while matrix factorization is adopted by GraphWave.

\subsubsection{Intra-community Proximity Preserving NRL}
Another interesting feature that real-world networks exhibit is the community structure, where vertices are densely connected to each other within the same community, but sparsely connected to vertices from other communities. For example, in social networks, people from the same interest group or affiliation often form a community. In citation networks, papers on similar research topics tend to frequently cite each other. Intra-community preserving NRL aims to leverage the community structure that characterizes key vertex properties to learn informative vertex representations.

\textbf{Learning Latent Social Dimensions}. The social dimension based NRL algorithms try to construct social actors' embeddings through their membership or affiliation to a number of social dimensions. To infer these latent social dimensions, the phenomenon of ``community" in social networks is considered, stating that social actors sharing similar properties often form groups with denser within-group connections. Thus, the problem boils down to one classical network analytic task---community detection---that aims to discover a set of communities with denser within-group connections than between-group connections. Three clustering techniques, including modularity maximization~\cite{tang2009relational}, spectral clustering~\cite{tang2011leveraging} and edge clustering~\cite{tang2009scalable} are employed to discover latent social dimensions. Each social dimension describes the likelihood of a vertex belonging to a plausible affiliation. These methods preserve the global community structure, but neglect local structure properties, \textit{e.g.}, the first-order and second-order proximity.

\textbf{Modularized Nonnegative Matrix Factorization (M-NMF)}. M-NMF~\cite{wang2017community} augments the second-order and high-order proximity with broader community structure to learn more informative vertex embeddings $U\in \mathbb{R}^{|V|\times d}$ using the following objective:
\begin{equation} \label{M_NMF_obj}
\small
\begin{aligned}
&\min_{M,U,H,C} \|S-MU^{\mathrm{T}}\|^{2}_{F}+\alpha\|H-UC^{\mathrm{T}}\|^{2}_{F}-\beta tr(H^{\mathrm{T}}BH)\\
&s.t.,\;\;M\geq 0,U\geq 0,H\geq 0,C\geq 0, tr(H^{\mathrm{T}}H)=|V|,
\end{aligned}
\end{equation}where vertex embedding $U$ is learned by minimizing $\|S-MU^{\mathrm{T}}\|^{2}_{F}$, with $S\in\mathbb{R}^{|V|\times|V|}$ being the vertex pairwise proximity matrix, which captures the second-order and the high-order proximity when taken as representations. The community indicative vertex embedding $H$ is learned by maximizing $tr(H^{\mathrm{T}}BH)$, which is essentially the objective of modularity maximization with $B$ being the modularity matrix. The minimization on $\|H-UC^{\mathrm{T}}\|^{2}_{F}$ makes these two embeddings consistent with each other by importing a community representation matrix $C$.

\textbf{Summary}: The algorithms of learning latent social dimensions ~\cite{tang2009relational,tang2011leveraging,tang2009scalable} only consider the community structure to learn vertex representation, while M-NMF~\cite{wang2017community} integrates microscopic structure (the second-order and high-order proximity) with the intra-community proximity. These methods primarily rely on matrix factorization to detect community structure, which makes them hard to scale up.

\subsubsection{Macroscopic Structure Preserving NRL}

Macroscopic structure preserving methods aim to preserve certain global network properties in a macroscopic view. Only very few recent studies are developed for this purpose.

%Existing research on structure preserving network representation learning mainly focuses on capturing the microscopic structure. As a result, the learned vertex representations are unable to represent the original networks in a macroscopic view. The macroscopic structure is therefore necessary to be considered for learning vertex representations. Some recent efforts have been made for macroscopic structure preserving network representation learning:

\textbf{Degree penalty principle (DP)}. Many real-world networks present the macroscopic scale-free property, which depicts the phenomenon that vertex degree follows a long-tailed distribution, i.e., most vertices are sparsely connected and only few vertices have dense edges. To capture the scale-free property, ~\cite{feng2018representation} proposes the degree penalty principle (DP): penalizing the proximity between high-degree vertices. This principle is then coupled with two NRL algorithms (i.e., spectral embedding~\cite{belkin2002laplacian} and DeepWalk~\cite{perozzi2014deepwalk}) to learn scale-free property preserving vertex representations.

\textbf{Hierarchical Representation Learning for Networks (HARP)}. To capture the global patterns in networks,  HARP~\cite{chen2018harp} samples small networks to approximate the global structure. The vertex representations learned from sampled networks are taken as the initialization for inferring the vertex representations of the original network. In this way, global structure is preserved in the final representations. To obtain smooth solutions, a series of smaller networks are successively sampled from the original network by coalescing edges and vertices, and the vertex representations are hierarchically inferred back from the smallest network to the original network. In HARP, DeepWalk~\cite{perozzi2014deepwalk} and LINE~\cite{tang2015line} are used to learn vertex representations.

\textbf{Summary}: DP~\cite{feng2018representation} and HARP~\cite{chen2018harp} are both realized by adapting the existing NRL algorithms to capture the macroscopic structure. The former tries to preserve the scale-free property, while the latter makes the learned vertex representations respect the global network structure. 

%\textbf{Summary}: Most of the above unsupervised structure preserving NRL algorithms are matrix factorization based methods and random walk based methods. The essence of matrix factorization based methods (i.e., Social dimensions~\cite{tang2009relational,tang2011leveraging,tang2009scalable}, GraRep~\cite{cao2015grarep}, HOPE~\cite{ou2016asymmetric}, M-NMF~\cite{wang2017community}) performing linear transformation on the original $|V|$-dimensional space that exhibits structure information, while deep learning based methods (DNGR~\cite{cao2016deep} and SDNE~\cite{wang2016structural}) use deep neural networks to realize non-linear transformation. These methods usually involve performing operation on the $|V|\times|V|$ scale matrix, making it hard to scale up. On the other hand, random walk based methods (i.e., DeepWalk~\cite{perozzi2014deepwalk}, node2vec~\cite{grover2016node2vec}, APP~\cite{zhou2017scalable}) exploits random walk to preserve local structure and they are easy to parallelize. 

\subsection{Unsupervised Content Augmented Network Representation Learning}
Besides network structure, real-world networks are often attached with rich content as vertex attributes, such as webpages in webpage networks, papers in citation networks, and user metadata in social networks. Vertex attributes provide direct evidence to measure content-level similarity between vertices. Therefore, network representation learning can be significantly improved if vertex attribute information is properly incorporated into the learning process. Recently, several content augmented NRL algorithms have been proposed to incorporate network structure and vertex attributes to reinforce the network representation learning. 

\subsubsection{Text-Associated DeepWalk (TADW)}

TADW~\cite{yang2015network} firstly proves the equivalence between DeepWalk~\cite{perozzi2014deepwalk} and the following matrix factorization:
\begin{equation}\label{TADW_obj1}\small
\min_{W,H}\|M-W^{\mathrm{T}}H\|^{2}_{F}+\frac{\lambda}{2}(\|W\|^{2}_{F}+\|H\|^{2}_{F}),
\end{equation}where $W$ and $H$ are learned latent embeddings and $M$ is the vertex-context matrix carrying transition probability between each vertex pair within $k$ steps. Then, textual features are imported through inductive matrix factorization \cite{natarajan2014inductive}
\begin{equation}\label{TADW_obj2}\small
\min_{W,H}\|M-W^{\mathrm{T}}HT\|^{2}_{F}+\frac{\lambda}{2}(\|W\|^{2}_{F}+\|H\|^{2}_{F}),
\end{equation}where $T$ is vertex textual feature matrix. After (\ref{TADW_obj2}) is solved, the final vertex representations are formed by taking the concatenation of $W$ and $HT$.  

\subsubsection{Homophily, Structure, and Content Augmented Network Representation Learning (HSCA)}

Despite its ability to incorporate textural features, TADW~\cite{yang2015network} only considers structural context of network vertices, i.e., the second-order and high-order proximity, but ignores the important homophily property (the first-order proximity) in its learning framework. HSCA~\cite{zhang2016homophily} is proposed to simultaneously integrates homophily, structural context, and vertex content to learn effective network representations.

For TADW, the learned representation for the $i$-th vertex $v_{i}$ is $\left[W_{:i}^{\mathrm{T}},(HT_{:i})^{\mathrm{T}}\right]^{\mathrm{T}}$, where $W_{:i}$ and $T_{:i}$ is the $i$-th column of $W$ and $T$, respectively. To enforce the first-order proximity, HSCA introduces a regularization term to enforce homophily between directly connected nodes in the embedding space, which is formulated as
\begin{equation}
\small
\mathcal{R}(W,H)=\frac{1}{4}\sum_{i,j=1}^{|V|}S_{ij}
\left\Vert \left[\begin{matrix}W_{:i}\\HT_{:i}\end{matrix}\right]-
\left[\begin{matrix}W_{:j}\\HT_{:j}\end{matrix}\right]\right\Vert_{2}^{2},
\end{equation}where $S$ is the adjacent matrix. The objective of HSCA is
\begin{equation}
\small
\min_{W,H}\|M-W^{\mathrm{T}}HT\|^{2}_{F}+\frac{\lambda}{2}(\|W\|^{2}_{F}+\|H\|^{2}_{F})+\mu\mathcal{R}(W,H),
\end{equation}where $\lambda$ and $\mu$ are the trade-off parameters. After solving the above optimization problem, the concatenation of $W$ and $HT$ is taken as the final vertex representations.

\subsubsection{Paired Restricted Boltzmann Machine (pRBM)}

By leveraging the strength of Restricted Boltzmann Machine (RBM)~\cite{hinton2006reducing}, \cite{wang2016paired} designs a novel model called Paired RBM (pRBM) to learn vertex representations by combining vertex attributes and link information. The pRBM considers the networks with vertices associated with binary attributes. For each edge $(v_{i},v_{j})\in E$, the attributes for $v_{i}$ and $v_{j}$ are $\bm{\mathrm{v}}^{(i)}$ and $\bm{\mathrm{v}}^{(j)}\in\{0,1\}^{m}$, and their hidden representations are $\bm{\mathrm{h}}^{(i)}$ and $\bm{\mathrm{h}}^{(j)}\in\{0,1\}^{d}$. Vertex hidden representations are learned by maximizing the joint probability of pRBM defined over $\bm{\mathrm{v}}^{(i)}$, $\bm{\mathrm{v}}^{(j)}$, $\bm{\mathrm{h}}^{(i)}$ and $\bm{\mathrm{h}}^{(j)}$:
\begin{equation} \label{pRBM_joint}
\small
\begin{aligned}
&\mathrm{Pr}(\bm{\mathrm{v}}^{(i)},\bm{\mathrm{v}}^{(j)},\bm{\mathrm{h}}^{(i)},\bm{\mathrm{h}}^{(j)},w_{ij};\bm{\theta})\\
&=\exp(-E(\bm{\mathrm{v}}^{(i)},\bm{\mathrm{v}}^{(j)},\bm{\mathrm{h}}^{(i)},\bm{\mathrm{h}}^{(j)},w_{ij}))/Z,
\end{aligned}
\end{equation}where $\bm{\theta}=\{\bm{\mathrm{W}}\in\mathbb{R}^{d\times m},\bm{\mathrm{b}}\in\mathbb{R}^{d\times 1},\bm{\mathrm{c}}\in\mathbb{R}^{m\times 1},\bm{\mathrm{M}}\in\mathbb{R}^{d\times d}\}$ is the parameter set and $Z$ is the normalization term.  To model the joint probability, the energy function is defined as
\begin{equation}
\small
\begin{aligned}
&E(\bm{\mathrm{v}}^{(i)},\bm{\mathrm{v}}^{(j)},\bm{\mathrm{h}}^{(i)},\bm{\mathrm{h}}^{(j)},w_{ij})=\\
&-w_{ij}(\bm{\mathrm{h}}^{(i)})^{\mathrm{T}}\bm{\mathrm{M}}\bm{\mathrm{h}}^{(j)}-(\bm{\mathrm{h}}^{(i)})^{\mathrm{T}}\bm{\mathrm{W}}\bm{\mathrm{v}}^{(i)}-\bm{\mathrm{c}}^{\mathrm{T}}\bm{\mathrm{v}}^{(i)}-\bm{\mathrm{b}}^{\mathrm{T}}\bm{\mathrm{h}}^{(i)}\\
&-(\bm{\mathrm{h}}^{(j)})^{\mathrm{T}}\bm{\mathrm{W}}\bm{\mathrm{v}}^{(j)}-\bm{\mathrm{c}}^{\mathrm{T}}\bm{\mathrm{v}}^{(i)}-\bm{\mathrm{b}}^{\mathrm{T}}\bm{\mathrm{h}}^{(j)},
\end{aligned}
\end{equation}where $w_{ij}(\bm{\mathrm{h}}^{(i)})^{\mathrm{T}}\bm{\mathrm{M}}\bm{\mathrm{h}}^{(j)}$ forces the latent representations of $v_{i}$ and $v_{j}$ to be close and $w_{ij}$ is the weight of edge $(v_{i},v_{j})$. 

\subsubsection{User Profile Preserving Social Network Embedding (UPP-SNE)}

UPP-SNE~\cite{zhang2017user} leverages user profile features to enhance the embedding learning of users in social networks. Compared with textural content features, user profiles have two unique properties: (1) user profiles are noisy, sparse and incomplete and (2) different dimensions of user profile features are topic-inconsistent. To filter out noise and extract useful information from user profiles, UPP-SNE constructs user representations by performing a non-linear mapping on user profile features, which is guided by network structure.

The approximated kernel mapping~\cite{rahimi2008random} is used in UPP-SNE to construct user embedding from user profile features:
\begin{equation}\label{mapping}
\begin{aligned}
\small
f(v_{i})=\varphi(\bm{x}_{i})=\frac{1}{\sqrt{d}}&\left[\cos(\bm{\mu}_{1}^{\mathrm{T}}\bm{x}_{i}),
\cdots,\cos(\bm{\mu}_{d}^{\mathrm{T}}\bm{x}_{i}),\right.\\
&\left.\sin(\bm{\mu}_{1}^{\mathrm{T}}\bm{x}_{i}),\cdots,\sin(\bm{\mu}_{d}^{\mathrm{T}}\bm{x}_{i})\right]^{\mathrm{T}},
\end{aligned}
\end{equation}where $\bm{x}_{i}$ is the user profile feature vector of vertex $v_{i}$ and $\bm{\mu}_{i}$ is the corresponding coefficient vector.

To supervise the learning of the non-linear mapping and make user profiles and network structure complement each other, the objective of DeepWalk~\cite{perozzi2014deepwalk} is used:
\begin{equation}
\small
\min_{f}\;-\log \mathrm{Pr}(\left\lbrace v_{i-t},\cdots, v_{i+t}\right\rbrace\setminus v_{i}\vert f(v_{i})),
\end{equation}where $\left\lbrace v_{i-t},\cdots, v_{i+t}\right\rbrace\setminus v_{i}$ is the context vertices of vertex $v_{i}$ within $t$ window size in the given random walk sequence.

\subsubsection{Property Preserving Network Embedding (PPNE)}

To learn content augmented vertex representations, PPNE~\cite{li2017ppne} jointly optimizes two objectives: (i) the structure-driven objective and (ii) the attribute-driven objective.

Following DeepWalk, the structure-driven objective aims to make vertices sharing similar context vertices represented closely. For a given random walk sequence $\mathcal{S}$, the structure-driven objective is formulated as
\begin{equation}\label{struct_obj}\small
\min D_{T}=\prod_{v\in\mathcal{S}}\prod_{u\in context(v)}\mathrm{Pr}(u|v).
\end{equation}The attribute-driven objective aims to make the vertex representations learned by Eq. (\ref{struct_obj}) respect the vertex attribute similarity. A realization of the attribute-driven objective is
\begin{equation}\small
\min D_{N}=\sum_{v\in\mathcal{S}}\sum_{u\in pos(v)\cup neg(v)}P(v,u)d(v,u),
\end{equation}where $P(u,v)$ is the attribute similarity between $u$ and $v$, $d(u,v)$ is the distance between $u$ and $v$ in the embedding space, and $pos(v)$ and $neg(v)$ is the set of top-$k$ similar and dissimilar vertices according to $P(u,v)$, respectively.

\textbf{Summary}: The above unsupervised content augmented NRL algorithms incorporate vertex content features in three ways. The first, used by TADW~\cite{yang2015network} and HSCA~\cite{zhang2016homophily}, is to couple the network structure with vertex content features via inductive matrix factorization~\cite{natarajan2014inductive}. This process can be considered as a linear transformation on vertex attributes constrained by network structure. The second is to perform a non-linear mapping to construct new vertex embeddings that respect network structure. For example, RBM~\cite{hinton2006reducing} and the approximated kernel mapping~\cite{rahimi2008random} is used by pRBM~\cite{wang2016paired} and UPP-SNE~\cite{zhang2017user}, respectively, to achieve this goal. The third used by PPNE~\cite{li2017ppne} is to add an attribute preserving constraint to the structure preserving optimization objective.

\section{Semi-supervised Network Representation Learning}
\label{sec:semi}

Label information attached with vertices directly indicates vertices' group or class affiliation. Such labels have strong correlations, although not always consistent, to network structure and vertex attributes, and are always helpful in learning informative and discriminative network representations. Semi-supervised NRL algorithms are developed along this line to make use of vertex labels available in the network for seeking more effective vertex representations.

\subsection{Semi-supervised Structure Preserving NRL}
The first group of semi-supervised NRL algorithms aim to simultaneously optimize the representation learning that preserves network structure and discriminative learning. As a result, the information derived from vertex labels can help improve the representative and discriminative power of the learned vertex representations.

\subsubsection{Discriminative Deep Random Walk (DDRW)}
Inspired by the discriminative representation learning~\cite{zhu2012medlda,mairal2009supervised}, DDRW~\cite{li2016discriminative} proposes to learn discriminative network representations through jointly optimizing the objective of DeepWalk ~\cite{perozzi2014deepwalk} together with the following L2-loss Support Vector Classification objective:
\begin{equation}\small
\mathcal{L}_{c}=C\sum_{i=1}^{|V|}(\sigma(1-Y_{ik}\beta^{\mathrm{T}}r_{v_{i}}))^{2}+\frac{1}{2}\beta^{\mathrm{T}}\beta,
\end{equation}where $\sigma(x)=x$, if $x>0$ and otherwise $\sigma(x)=0$.

The joint objective of DDRW is thus defined as
\begin{equation} \label{DDRW_obj}\small
\mathcal{L}=\eta\mathcal{L}_{DW}+\mathcal{L}_{c}.
\end{equation}where $\mathcal{L}_{DW}$ is the objective function of Deekwalk. The objective (\ref{DDRW_obj}) aims to learn discriminative vertex representations for binary classification for the $k$-th class. DDRW is generalized to handle multi-class classification by using the \textit{one-against-rest} strategy~\cite{fan2008liblinear}.

\subsubsection{Max-Margin DeepWalk (MMDW)}

Similarly, MMDW~\cite{tu2016max} couples the objective of the matrix factorization version DeepWalk~\cite{yang2015network} with the following multi-class Support Vector Machine objective with $\{(r_{v_{1}},Y_{1:}),\cdots,(r_{v_{T}},Y_{T:})\}$ training set:
\begin{equation}\label{MMDW_obj2}
\begin{aligned}\small
&\min_{W,\xi}\mathcal{L}_{SVM}=\min_{W,\xi}\frac{1}{2}\|W\|^{2}_{2}+C\sum_{i=1}^{T}\xi_{i},\\
&s.t.\;\mathrm{w}_{l_{i}}^{\mathrm{T}}r_{v_{i}}-\mathrm{w}_{j}^{\mathrm{T}}r_{v_{i}}\geq e^{j}_{i}-\xi_{i},\; \forall i,j,
\end{aligned}
\end{equation}where $l_{i}=k$ with $Y_{ik}=1$, $e^{j}_{i} = 1$ for $Y_{ij}=-1$, and $e^{j}_{i} = 0$ for $Y_{ij}=1$.
%\begin{equation}
%e^{j}_{i}=\left\{
%\begin{aligned}
%&1,\;\mathrm{if}\;Y_{ij}=-1,\\
%&0,\;\mathrm{if}\;Y_{ij}=1.
%\end{aligned}\right.
%\end{equation}Here, $W=[\mathrm{w}_{1},\cdots,\mathrm{w}_{m}]^{\mathrm{T}}$ is the weight matrix of SVM, $l_{i}=j$ for $Y_{ij}=1$ and $\xi=[\xi_{1},\cdots,\xi_{T}]$ is the slack variable that tolerates classification errors in the training set.

The joint objective of MMDW is
\begin{equation}\label{MMDW_obj}
\begin{aligned}\small
&\min_{U,H,W,\xi}\mathcal{L}=\min_{U,H,W,\xi}\mathcal{L}_{DW}+\frac{1}{2}\|W\|^{2}_{2}+C\sum_{i=1}^{T}\xi_{i},\\
&s.t.\;\mathrm{w}_{l_{i}}^{\mathrm{T}}r_{v_{i}}-\mathrm{w}_{j}^{\mathrm{T}}r_{v_{i}}\geq e^{j}_{i}-\xi_{i},\; \forall i,j.
\end{aligned}
\end{equation}where $\mathcal{L}_{DW}$ is the objective of the matrix factorization version of DeepWalk.

\subsubsection{Transductive LINE (TLINE)}
Along similar lines, TLINE~\cite{zhang2016tline} is proposed as a semi-supervised extension of LINE~\cite{tang2015line} that simultaneously learns LINE's vertex representations and an SVM classifier. Given a set of labeled and unlabeled vertices $\{v_{1},v_{2},\cdots,v_{L}\}$ and $\{v_{L+1},\cdots,v_{|V|}\}$, TLINE trains a multi-class SVM classifier on $\{v_{1},v_{2},\cdots,v_{L}\}$ by optimizing the objective:
\begin{equation}\label{tline_svm}\small
\mathcal{O}_{svm}=\sum_{i=1}^{L}\sum_{k=1}^{K}\mathrm{max}(0,1-Y_{ik}{\mathrm{w}_{k}}^{\mathrm{T}}r_{v_{i}})+\lambda\Vert \mathrm{w}_{k}\Vert_{2}^{2}.
\end{equation}

Based on LINE's formulations that preserve the first-order and second-order proximity, TLINE optimizes two objective functions:
\begin{equation}\small
\mathcal{O}_{TLINE(1st)}=\mathcal{O}_{line1}+\beta\mathcal{O}_{svm},
\end{equation}
%where $\beta$ is the trade-off parameter. Similarly, the second-order proximity is also exploited in TLINE to learn discriminative vertex representations by minimizing the following objective:
\begin{equation}\small
\mathcal{O}_{TLINE(2nd)}=\mathcal{O}_{line2}+\beta\mathcal{O}_{svm}.
\end{equation}
Inheriting LINE's ability to deal with large-scale networks, TLINE is claimed to be able to learn discriminative vertex representations for large-scale networks with low time and memory cost.

\subsubsection{Group Enhanced Network Embedding (GENE)}
GENE~\cite{chen2016incorporate} integrates group (label) information with network structure in a probabilistic manner. GENE assumes that vertices should be embedded closely in low-dimensional space, if they share similar neighbors or join similar groups. Inspired by DeepWalk \cite{perozzi2014deepwalk} and document modeling \cite{le2014distributed,djuric2015hierarchical}, the mechanism of GENE for learning group label informed vertex representations is achieved by maximizing the following log probability:
\begin{equation}
\begin{small}
\begin{aligned}
\mathscr{L}=\sum_{g_{i}\in \mathcal{Y}}&\left[\alpha\sum_{W\in W_{g_{i}}}\sum_{v_{j}\in W}\log\mathrm{Pr}(v_{j}|v_{j-t},\cdots,v_{j+t},g_{i})+\right.\\
&\left.\beta\sum_{\hat{v}_{j}\in\hat{W}_{g_{j}}}\log\mathrm{Pr}(\hat{v}_{j}|g_{i})\right],
\end{aligned}
\end{small}
\end{equation}where $\mathcal{Y}$ is the set of different groups, $W_{g_{i}}$ is the set of random walk sequences labeled with $g_{i}$, $\hat{W}_{g_{i}}$ is the set of vertices randomly sampled from group $g_{i}$. 

\subsubsection{Semi-supervised Network Embedding (SemiNE)}
SemiNE~\cite{li2017semi} learns semi-supervised vertex representations in two stages. In the first stage, SemiNE exploits the DeepWalk~\cite{perozzi2014deepwalk} framework to learn vertex representations in an unsupervised manner. It points out that DeepWalk does not consider the order information of context vertex, i.e., the distance between the context vertex and the central vertex, when using the context vertex $v_{i+j}$ to predict the central vertex $v_{i}$. Thus, SemiNE encodes the order information into DeepWalk by modeling the probability $\mathrm{Pr}(v_{i+j}|v_{i})$ with $j$-dependent parameters:
\begin{equation}\small
\mathrm{Pr}(v_{i+j}|v_{i})=\frac{\exp(\Phi(v_{i})\cdot\Psi_{j}(v_{i+j}))}{\sum_{u\in V}\exp(\Phi(v_{i})\cdot\Psi_{j}(u))},
\end{equation}where $\Phi(\cdot)$ is the vertex representation and $\Psi_{j}(\cdot)$ is the parameter for calculating $\mathrm{Pr}(v_{i+j}|v_{i})$.

In the second stage, SemiNE learns a neural network that tunes the learned unsupervised vertex representations to fit vertex labels.

%\textbf{Summary}: In general, three strategies are adopted by the existing semi-supervised structure preserving NRL algorithms to empower vertex labels: the first (i.e., DDRW~\cite{li2016discriminative}, MMDW~\cite{tu2016max} and TLINE~\cite{zhang2016tline}) is to enforce a classification loss on the objective function of representation learning, the second (used by GENE~\cite{chen2016incorporate}) is to maximize the conditional probability of co-occurring vertices given vertex labels, and the third (used by SemiNE~\cite{li2017semi} is to tune the learned unsupervised vertex representations to fit vertex labels. Among others, MMDW~\cite{tu2016max} suffers from the same scalability problem as other matrix factorization based methods. 

\subsection{Semi-supervised Content Augmented NRL}
Recently, more research efforts have shifted to the development of label and content augmented NRL algorithms that investigate the use of vertex content and labels to assist with network representation learning. With content information incorporated, the learned vertex representations are expected to be more informative, and with label information considered, the learned vertex representations can be highly customized for the underlying classification task.

\subsubsection{Tri-Party Deep Network Representation (TriDNR)}
Using a coupled neural network framework, TriDNR~\cite{pan2016tri} learns vertex representations from three information sources: network structure, vertex content and vertex labels. To capture the vertex content and label information, TriDNR adapts the Paragraph Vector model ~\cite{le2014distributed} to describe the vertex-word correlation and the label-word correspondence by maximizing the following objective:
\begin{equation}\label{TriDNR_obj2}\small
\mathcal{L}_{PV}=\sum_{i\in L}\log\mathrm{Pr}(w_{-b}:w_{b}|c_{i})+\sum_{i=1}^{|V|}\log\mathrm{Pr}(w_{-b}:w_{b}|v_{i}),
\end{equation}where $\{w_{-b}:w_{b}\}$ is a sequence of words inside a contextual window of length $2b$, $c_{i}$ is the class label of vertex $v_{i}$, and $L$ is the set of indices of labeled vertices.

TriDNR is then realized by coupling the Paragraph Vector objective with DeepWalk objective:
\begin{equation}\small
\max\;(1-\alpha)\mathcal{L}_{DW}+\alpha\mathcal{L}_{PV},
\end{equation}where $\mathcal{L}_{DW}$ is the DeepWalk maximization objective function and $\alpha$ is the trade-off parameter.

\subsubsection{Linked Document Embedding (LDE)}
LDE~\cite{wang2016linked} is proposed to learn representations for linked documents, which are actually the vertices of citation or webpage networks. Similar to TriDNR~\cite{pan2016tri}, LDE learns vertex representations by modeling three kinds of relations, i.e., word-word-document relations, document-document relations, and document-label relations. LDE is realized by solving the following optimization problem:
\begin{equation}\label{LDE_obj}
\small
\begin{aligned}
\min_{\bm{W},\bm{D},\bm{Y}}&-\frac{1}{|\mathcal{P}|}\sum_{(w_{i},w_{j},d_{k})\in\mathcal{P}}\log\mathrm{Pr}(w_{j}|w_{i},d_{k})\\
&-\frac{1}{|E|}\sum_{i}\sum_{j:(v_{i},v_{j})\in E}\log\mathrm{Pr}(d_{j}|d_{i})\\
&-\frac{1}{|\mathcal{Y}|}\sum_{i:y_{i}\in\mathcal{Y}}\log\mathrm{Pr}(y_{i}|d_{i})\\
&+\gamma(\|\bm{W}\|^{2}_{F}+\|\bm{D}\|^{2}_{F}+\|\bm{Y}\|^{2}_{F}).
\end{aligned}
\end{equation}Here, the probability $\mathrm{Pr}(w_{j}|w_{i},d_{k})$ is used to model word-word-document relations, which means the probability that in document $d_{k}$, word $w_{j}$ is a neighboring word of $w_{i}$. To capture word-word-document relations, triplets $(w_{i},w_{j},d_{k})$ are extracted, with the word-neighbor pair $(w_{i},w_{j})$ occurring in document $d_{k}$. The set of triplets $(w_{i},w_{j},d_{k})$ is denoted by $\mathcal{P}$. The document-document relations are captured by the conditional probability between linked document pairs $(d_{i},d_{j})$, $\mathrm{Pr}(d_{j}|d_{i})$. The document-label relations are also considered by modeling $\mathrm{Pr}(y_{i}|d_{i})$, the probability for the occurrence of class label $y_{i}$ conditioned on document $d_{i}$. In (\ref{LDE_obj}), $\bm{W}$, $\bm{D}$ and $\bm{Y}$ is the embedding matrix for words, documents and labels, respectively. %To avoid overfitting, a regularization term $\|\bm{W}\|^{2}_{F}+\|\bm{D}\|^{2}_{F}+\|\bm{Y}\|^{2}_{F}$ is added.  

\subsubsection{Discriminative Matrix Factorization (DMF)}
To empower vertex representations with discriminative ability, DMF~\cite{zhang2016collective} enforces the objective of TADW (\ref{TADW_obj2}) with an empirical loss minimization for a linear classifier trained on labeled vertices:
\begin{equation}\label{DMF_obj3}
\small
\begin{aligned}
\min_{W,H,\bm{\eta}}&\frac{1}{2}\sum_{i,j=1}^{|V|}(M_{ij}-\bm{w}^{\mathrm{T}}_{i}H\bm{t}_{j})^{2}+\frac{\mu}{2}\sum_{n\in\mathcal{L}}(Y_{n1}-\bm{\eta}^{\mathrm{T}}\bm{x}_{n})^{2}\\
&+\frac{\lambda_{1}}{2}(\|H\|_{F}^{2}+\|\bm{\eta}\|_{2}^{2})+\frac{\lambda_{2}}{2}\|W\|_{F}^{2},
\end{aligned}
\end{equation}where $\bm{w}_{i}$ is the $i$-th column of vertex representation matrix $W$ and $\bm{t}_{j}$ is $j$-th column of vertex textual feature matrix $T$, and $\mathcal{L}$ is the set of indices of labeled vertices. DMF considers binary-class classification, i.e. $\mathcal{Y}=\{+1,-1\}$. Hence, $Y_{n1}$ is used to denote the class label of vertex $v_{n}$. 

DMF constructs vertex representations from $W$ rather that $W$ and $HT$. This is based on empirical findings that $W$ contains sufficient information for vertex representations. In the objective of (\ref{DMF_obj3}), $\bm{x}_{n}$ is set to $[\bm{w}_{n}^{\mathrm{T}},1]^{\mathrm{T}}$, which incorporates the intercept term $b$ of the linear classifier into $\bm{\eta}$. The optimization problem (\ref{DMF_obj3}) is solved by optimizing $W$, $H$ and $\bm{\eta}$ alternately. Once the optimization problem is solved, the discriminative and informative vertex representations together with the linear classifier are learned, and work together to classify unlabeled vertices in networks.

\subsubsection{Predictive Labels And Neighbors with Embeddings Transductively Or Inductively from Data (Planetoid)} Planetoid~\cite{yang2016revisiting} leverages network embedding together with vertex attributes to carry out semi-supervised learning. Planetoid learns vertex embeddings by minimizing the loss for predicting structural context, which is formulated as
\begin{equation}\label{Planetoid_obj1}\small
\mathcal{L}_{u}=-\mathbb{E}_{(i,c,\gamma)}\log\sigma(\gamma\mathrm{w}_{c}^{\mathrm{T}}\bm{\mathrm{e}}_{i}),
\end{equation}where $(i, c)$ is the index for vertex context pair $(v_{i},v_{c})$, $\bm{\mathrm{e}}_{i}$ is the embedding of vertex $v_{i}$, $\mathrm{w}_{c}$ is the parameter vector for context vertex $v_{c}$, and $\gamma\in\{+1,-1\}$ indicates whether the sampled vertex context pair $(i,c)$ is positive or negative. The triple $(i,c,\gamma)$ is sampled according to both the network structure and vertex labels.%, which encodes the information from these two sources into vertex representation $\bm{\mathrm{e}_{i}}$.

Planetoid then maps the learned vertex representations $\bm{\mathrm{e}}$ and vertex attributes $\bm{\mathrm{x}}$ to hidden layer space via deep neural network, and concatenates these two hidden layer representations together to predict vertex labels, by minimizing the following classification loss:
\begin{equation}\label{Planetoid_obj2}\small
\mathcal{L}_{s}=-\frac{1}{L}\sum_{i=1}^{L}\log p(y_{i}|\bm{\mathrm{x}}_{i},\bm{\mathrm{e}_{i}}),
\end{equation}
%where
%\begin{equation}
%p(y_{i}|\mathrm{x}_{i},\mathrm{e}_{i})=\frac{\exp\left(\left[\mathrm{h}^{k}(\bm{\mathrm{x}}_{i})^{\mathrm{T}},\mathrm{h}^{l}(\bm{\mathrm{e}}_{i})^{\mathrm{T}}\right]\cdot\mathrm{w}_{y_{i}}\right)}
%{\sum_{y^{\prime}}\exp\left(\left[\mathrm{h}^{k}(\bm{\mathrm{x}}_{i})^{\mathrm{T}},\mathrm{h}^{l}(\bm{\mathrm{e}}_{i})^{\mathrm{T}}\right]\cdot\mathrm{w}_{y^{\prime}}\right)}.
%\end{equation}Here, $\mathrm{h}^{k}(\cdot)$ and $\mathrm{h}^{l}(\cdot)$ is the deep neural network transformation for vertex attributes $\bm{\mathrm{x}}$ and vertex embedding $\bm{\mathrm{e}}$, and $\mathrm{w}_{y}$ is the parameter vector for class $y$.

To integrate network structure, vertex attributes and vertex labels together, Planetoid jointly minimizes the two objectives (\ref{Planetoid_obj1}) and (\ref{Planetoid_obj2}) to learn vertex embedding $\bm{\mathrm{e}}$ with deep neural networks.

%Planetoid is then extended to inductive setting via constructing vertex embedding $\bm{\mathrm{e}}$ from vertex attributes $\bm{\mathrm{x}}$ with deep neural network.

\subsubsection{Label informed Attribute Network Embedding (LANE)}
LANE~\cite{huang2017label} learns vertex representations by embedding the network structure proximity, attribute affinity, and label proximity into a unified latent representation. The learned representations are expected to capture both network structure and vertex attribute information, and label information if provided. The embedding learning in LANE is carried out in two stages. During the first stage, vertex proximity in network structure and attribute information are mapped into latent representations $\bm{\mathrm{U}}^{(G)}$ and $\bm{\mathrm{U}}^{(A)}$, then $\bm{\mathrm{U}}^{(A)}$ is incorporated into $\bm{\mathrm{U}}^{(G)}$ by maximizing their correlations. In the second stage, LANE employs the joint proximity (determined by $\bm{\mathrm{U}}^{(G)}$) to smooth label information and uniformly embeds them into another latent representation $\bm{\mathrm{U}}^{(Y)}$, and then embeds $\bm{\mathrm{U}}^{(A)}$, $\bm{\mathrm{U}}^{(G)}$ and $\bm{\mathrm{U}}^{(Y)}$ into a unified embedding representation $\bm{\mathrm{H}}$. 

%\textbf{Summary}: The above semi-supervised content augmented NRL algorithms use three strategies to fully integrate network structure, vertex content and vertex labels to seek more informative and discriminative vertex representations. The first, used by TriDNR~\cite{pan2016tri} and LDE~\cite{wang2016linked}is to maximize the conditional co-occurrence probability of vertex-word relation, vertex-vertex structure relation and vertex-label relation. The second used by DMF~\cite{zhang2016collective} and Planetoid~\cite{yang2016revisiting} is to enforce the content augmented vertex representations with a classification objective. The third used by LANE~\cite{huang2017label} is to unify vertex structure latent representations, vertex content latent representations and vertex label latent representations via embedding them into a joint space.  

\subsection{Summary}

We now summarize and compare the discriminative learning strategies used by semi-supervised NRL algorithms in Table~\ref{table-semiNRL} in terms of their advantages and disadvantages.

	\begin{table*}[!htbp]
		\centering
		\scriptsize
		\tabcolsep 3pt
		\renewcommand\arraystretch{1.3}
		\caption{A summary of semi-supervised NRL algorithms}
		\begin{tabular}{|c|c|c|c|c|}
			\hline
			\bf Discriminative Learning Strategy & \bf Algorithm &  \bf Loss function & \bf Advantage & \bf Disadvantage \\\hline
			\multirow{5}{*}{fitting a classifier} &  DDRW~\cite{li2016discriminative} & hinge loss & \multirow{5}{*}{\tabincell{c}{a) directly optimize classification loss;\\ b) perform better in sparsely labeled scenarios} } & \multirow{5}{*}{prone to overfitting}\\\cline{2-3}
			& MMDW~\cite{tu2016max} & hinge loss & &\\\cline{2-3}
			& TLINE~\cite{zhang2016tline} & hinge loss & &\\\cline{2-3}
			& DMF~\cite{zhang2016collective} & square loss & &\\\cline{2-3}
			& SemiNE~\cite{li2017semi} & logistic loss & &\\\hline
			\multirow{4}{*}{modeling vertex label relation} & GENE~\cite{chen2016incorporate} & likelihood loss & \multirow{5}{*}{\tabincell{c}{a) better capture intra-class proximity;\\ b) generalization to other tasks}} & \multirow{5}{*}{require more labeled data} \\\cline{2-3}
			& TriDNR~\cite{pan2016tri} & likelihood loss & &\\\cline{2-3}
			& LDE~\cite{wang2016linked} & likelihood loss & &\\\cline{2-3}
			& Planetoid~\cite{yang2016revisiting} & likelihood loss & &\\\cline{1-3}
			joint vertex label embedding & LANE~\cite{huang2017label} & correlation loss & &\\\hline
		\end{tabular}
		\label{table-semiNRL}
	\end{table*}

Three strategies are used to achieve discriminative learning. The first strategy (\textit{i.e.}, DDRW~\cite{li2016discriminative}, MMDW~\cite{tu2016max}, TLINE~\cite{zhang2016tline}, DMF~\cite{zhang2016collective}, SemiNE~\cite{li2017semi}) is to enforce classification loss minimization on vertex representations, \textit{i.e.}, fitting the vertex representations to a classifier. This provides a direct way to separate vertices of different categories from each other in the new embedding space. The second strategy (used by GENE~\cite{chen2016incorporate}, TriDNR~\cite{pan2016tri}, LDE~\cite{wang2016linked} and Planetoid~\cite{yang2016revisiting}) is achieved by modeling vertex label relation, such that vertices with same labels have similar vector representations. The third strategy used by LANE~\cite{huang2017label} is to jointly embed vertices and labels into a common space. 

Fitting vertex representations to a classifier can take advantage of the discriminative power in vertex labels. Algorithms using this strategy only require a small number of labeled vertices (\textit{e.g.}, 10\%) to achieve significant performance gain over their unsupervised counterparts. They are thus more effective for discriminative learning in sparsely labeled scenarios. However, fitting vertex representations to a classifier is more prone to overfitting. Regularization and DropOut~\cite{hinton2012improving} are often introduced to overcome this problem. By contrast, modeling vertex label relation and joint vertex embedding requires more vertex labels to make vertex representations more discriminative, but they can better capture intra-class proximity, \textit{i.e.}, vertices belonging to the same class are kept closer to each other in the new embedding space. This allows them to have generalized benefits on tasks like vertex clustering or visualization. 

\section{Applications}
\label{sec:app}

%Network representation learning techniques have been successfully applied to undertake a series of network analytic tasks across many disciplines. 

Once new vertex representations are learned via network representation learning techniques, traditional vector-based algorithms can be used to solve important analytic tasks, such as vertex classification, link prediction, clustering, visualization, and recommendation. The effectiveness of the learned representations can also be validated through assessing their performance on these tasks.

\subsection{Vertex Classification}

Vertex classification is one of the most important tasks in network analytic research. Often in networks, vertices are associated with semantic labels characterizing certain aspects of entities, such as beliefs, interests, or affiliations. In citation networks, a publication may be labeled with topics or research areas, while the labels of entities in social network may indicate individuals' interests or political beliefs. Often, because network vertices are partially or sparsely labeled due to high labeling costs, a large portion of vertices in networks have unknown labels. The problem of vertex classification aims to predict the labels of unlabeled vertices given a partially labeled network~\cite{zhu2007combining,bhagat2011node}. Since vertices are not independent but connected to each other in the form of a network via links, vertex classification  should exploit these dependencies for jointly classifying the labels of vertices. Among others, collective classification proposes to construct a new set of vertex features that summarize label dependencies in the neighborhood, which has been shown to be most effective in classifying many real-world networks~\cite{sen2008collective,kazienko2012label}.

Network representation learning follows the same principle that automatically learns vertex features based on network structure. Existing studies have evaluated the discriminative power of the learned vertex representations under two settings: unsupervised settings (\textit{e.g.}, \cite{perozzi2014deepwalk,zhang2016homophily,tang2015line,yang2015network,grover2016node2vec}), where vertex representations are learned separately, followed by applying discriminative classifiers like SVM or logistic regression on the new embeddings, and semi-supervised settings (\textit{e.g.}, ~\cite{li2016discriminative,zhang2016collective,tu2016max,zhang2016tline,huang2017label}), where representation learning and discriminative learning are simultaneously tackled, so that discriminative power inferred from labeled vertices can directly benefit the learning of informative vertex representations. These studies have proved that better vertex representations can contribute to high classification accuracy.

\subsection{Link Prediction}

Another important application of network representation learning is link prediction~\cite{liben2007link,gao2011temporal}, which aims to infer the existence of new relationships or emerging interactions between pairs of entities based on the currently observed links and their properties. The approaches developed to solve this problem can enable the discovery of implicit or missing interactions in the network, the identification of spurious links, as well as understanding the network evolution mechanism. Link prediction techniques are widely applied in social networks to predict unknown connections among people, which can be used to recommend friendship or identify suspicious relationships. Most of the current social networking systems are using link prediction to automatically suggest friends with a high degree of accuracy. In biological networks, link prediction methods have been developed to predict previously unknown interactions between proteins, thus significantly reducing the costs of empirical approaches. Readers can refer to the survey papers~\cite{lu2011link,martinez2017survey} for the recent progress in this field.

Good network representations should be able to capture explicit and implicit connections between network vertices thus enabling application to link prediction.  \cite{wang2016structural} and \cite{ou2016asymmetric} predict missing links based on the learned vertex representations on social networks. \cite{grover2016node2vec} also applies network representation learning to collaboration networks and protein-protein interaction networks. They demonstrate that on these networks links predicted using the learned representations achieve better performance than traditional similarity-based link prediction approaches.

\subsection{Clustering}

Network clustering refers to the task of partitioning network vertices into a set of clusters, such that vertices are densely connected to each other within the same cluster, but connected to few vertices from other clusters~\cite{fortunato2010community}. Such cluster structures, or communities widely occur in a wide spectrum of networked systems from bioinformatics, computer science, physics, sociology, \textit{etc.}, and have strong implications. For example, in biology networks, clusters may correspond to a group of proteins having the same function; in the network of webpages, clusters are likely pages having similar topics or related content; in social networks, clusters may indicate groups of people having similar interests or affiliations.

Researchers have proposed a large body of network clustering algorithms based on various metrics of similarity or strength of connection between vertices. Min-max cut and normalized cut methods~\cite{ding2001minmax,shi2000normalized} seek to recursively partition a graph into two clusters that maximize the number of intra-cluster connections and minimize the number of inter-cluster connections. Modularity-based methods (\textit{e.g.}, \cite{newman2004finding,newman2006modularity}) aim to maximize the modularity of a clustering, which is the fraction of intra-cluster edges minus the expected fraction assuming the edges were randomly distributed. A network partitioning with high modularity would have dense intra-cluster connections but sparse inter-cluster connections. Some other methods (\textit{e.g.}, \cite{xu2007scan}) try to identify nodes with similar structural roles like bridges and outliers.

Recent NRL methods (\textit{e.g.}, GraRep~\cite{cao2015grarep}, DNGR~\cite{cao2016deep}, M-NMF~\cite{wang2017community}, and pRBM~\cite{wang2016paired}) used the clustering performance to evaluate the quality of the learned network representations on different networks. Intuitively, better representations would lead to better clustering performance. These works followed the common approach that first applies an unsupervised NRL algorithm to learn vertex representations, and then performs $k$-means clustering on the learned representations to cluster the vertices. In particular, pRBM~\cite{wang2016paired} showed that NRL methods outperform the baseline that uses original features for clustering without learning representations. This suggests that effective representation learning can improve the clustering performance.

\subsection{Visualization}

Visualization techniques play critical roles in managing, exploring, and analyzing complex networked data. \cite{herman2000graph} surveys a range of methods used to visualize graphs from an information visualization perspective. This work compares various traditional layouts used to visualize graphs, such as tree-, 3D-, and hyperbolic-based methods, and shows that classical visualization techniques are proved effective for small or intermediate sized networks; they however confront a big challenge when applied to large-scale networks. Few systems can claim to deal effectively with thousands of vertices, although networks with this order of magnitude often occur in a wide variety of applications. Consequently, a first step in the visualization process is often to reduce the size of the network to display. One common approach is essentially to find an extremely low-dimensional representation of a network that preserves the intrinsic structure, i.e., keeping similar vertices close and dissimilar vertices far apart, in the low-dimensional space~\cite{tang2016visualizing}.

Network representation learning has the same objective that embeds a large network into a new latent space of low dimensionality. After new embeddings are obtained in the vector space, popular methods such as $t$-distributed stochastic neighbor embedding ($t$-SNE)~\cite{maaten2008visualizing} can be applied to visualize the network in a 2-D or 3-D space. By taking the learned vertex representations as input, LINE~\cite{tang2015line} used the $t$-SNE package to visualize the DBLP co-author network after the authors are mapped into a 2-D space, and showed that LINE is able to cluster authors in the same field to the same community. HSCA~\cite{zhang2016homophily} illustrated the advantages of the content-augmented NRL algorithm by visualizing citations networks. Semi-supervised algorithms (\textit{e.g.}, TLINE~\cite{zhang2016tline}, TriDNR~\cite{pan2016tri}, and DMF~\cite{zhang2016collective}) demonstrated that the visualization results have better clustering structures with vertex labels properly imported.

\subsection{Recommendation}

In addition to structure, content, and vertex label information, many social networks also include geographical and spatial-temporal information, and users can share their experiences online with their friends for point of interest (POI) recommendation, \textit{e.g.}, transportation, restaurant, and sightseeing landmark, \textit{etc.} Examples of such location-based social networks (LBSN) include Foursquare, Yelp, Facebook Places, and many others. For these types of social networks, POI recommendation intends to recommend user interested objects, depending on their own context, such as the geographic location of the users and their interests. Traditionally, this is solved by using approaches, such as collaborative filtering, to leverage spatial and temporal correlation between user activities and geographical distance~\cite{zheng2009mining}. However, because each user's check-in records are very sparse, finding similar users or calculating transition probability between users and locations is a significant challenge. 

Recently, spatial-temporal embedding~\cite{xie2016learning,zhang2017regions,wang2018ensemble} has emerged to learn low-dimensional dense vectors to represent users, locations, and point-of-interests \textit{etc.} As a result, each user, location, and POI can be represented as a low-dimensional vector, respectively, for similarity search and many other analysis. An inherent advantage of such spatial-temporal aware embedding is that it alleviates the data sparsity problem, because the learned low dimensional vector is typically much more dense than the original representation. As a result, it makes query tasks, such as top-$k$ POI search, much more accurate than traditional approaches.

\subsection{Knowledge Graph}

Knowledge graphs represent a new type of data structure in database systems which encode structured information of billions of entities and their rich relations. A knowledge graph typically contains a rich set of heterogeneous objects and different types of entity relationships. Such networked entities form a gigantic graph and is now powering many commercial search engines to find similar objects online. Traditionally, knowledge graph search is carried out through database driven approaches to explore schema mapping between entities, including entity relationships. Recent advancement in network representation learning has inspired structured embeddings of knowledge bases~\cite{bordes2011learning}. Such embedding methods learn a low-dimensional vector representation for knowledge graph entities, such that generic database queries, such as top-$k$ search, can be carried out by comparing vector representation of the query object and objects in the database.

In addition to using vector representation to represent knowledge graph entities, researchers have also proposed to use such representation to further enhance and complete the knowledge graph itself. For example, knowledge graph completion intends to discover complete relationships between entities, and a recent work~\cite{feng2016gake} has proposed to use graph context to find missing links between entities. This is similar to link prediction in social networks, but the entities are typically heterogeneous and a pair of entities may also have different types of relationships.

\section{Evaluation Protocols}
\label{sec:eva}

%In order to make fair comparisons of the effectiveness of different NRL algorithms, a broad range of real-world information networks have been used in the literature for analysis and evaluation. 

In this section, we discuss evaluation protocols for validating the effectiveness of network representation learning. This includes a summary of commonly used benchmark datasets and evaluation methods, followed by a comparison of algorithm performance and complexity.

\subsection{Benchmark Datasets}

\begin{table*}[h]
	\tabcolsep 5pt
	\scriptsize
	\renewcommand\arraystretch{1.3}
	\caption{A summary of benchmark datasets for evaluating network representation learning.}
	\centering
	\scalebox{0.92}{
		\begin{minipage}{\linewidth}
			\centering
			\begin{tabular}{|c|c|c|c|c|c|c|c|}
				\hline
				\bf Category &\bf Dataset & Type & \bf $|V|$ & \bf $|E|$ & \bf $|\mathcal{Y}|$ & \bf Multi-label & \bf Vertex attr.\\\hline
				\multirow{8}{*}{\bf Social Network} & BlogCatalog\footnote{\url{http://www.public.asu.edu/\~ltang9/}} & undirected, binary & 10,312 & 333,983 & 39 & Yes & No \\\cline{2-8}
				& Flickr\footnote{\url{http://socialnetworks.mpi-sws.org/data-imc2007.html}\label{fn:Flickr}} & undirected, binary & 80,513 & 5,899,882 & 195 & Yes & No \\\cline{2-8}
				& YouTube\footref{fn:Flickr} & undirected, binary & 1,138,499 & 2,990,443 & 47 & Yes & No \\\cline{2-8}
				& Facebook\footnote{\url{https://snap.stanford.edu/data/egonets-Facebook.html}} & undirected, binary & 4,039 & 88,234 & 4 & No & Yes \\\cline{2-8}
				& Amherst\footnote{\url{https://escience.rpi.edu/data/DA/fb100/}\label{fn:Amherst}} \cite{traud2012social} & undirected, binary & 2,021 & 81,492 & 15 & No & Yes \\\cline{2-8}
				& Hamilton\footref{fn:Amherst} \cite{traud2012social} & undirected, binary & 2,118 & 87,486 & 15 & No & Yes \\\cline{2-8}
				& Mich\footref{fn:Amherst} \cite{traud2012social} & undirected, binary & 2,933 & 54,903 & 13 & No & Yes \\\cline{2-8}
				& Rochester\footref{fn:Amherst} \cite{traud2012social} & undirected, binary & 4,145 & 145,305 & 19 & No & Yes \\\hline
				\bf Language Network & Wikipedia \cite{tang2015line} & undirected, weighted & 1,985,098 & 1,000,924,086 & N/A & N/A & No\\\hline
				\multirow{6}{*}{\bf Citation Network} & DBLP (PaperCitation)\cite{tang2015line,tang2008arnetminer}& directed, binary & 781,109 & 4,191,677 & 7 & No & Yes \\\cline{2-8}
				& DBLP (AuthorCitation)\cite{tang2015line,tang2008arnetminer}& directed, weighted & 524,061 & 20,580,238 & 7 & No & No \\\cline{2-8}
				& Cora\footnote{\url{https://linqs.soe.ucsc.edu/data}\label{fn:Cora}} & directed, binary & 2,708 & 5,429 & 7 & No & Yes \\\cline{2-8}
				& Citeseer\footref{fn:Cora} & directed, binary & 3,312 & 4,732 & 6 & No & Yes \\\cline{2-8}
				& PubMed\footref{fn:Cora} & directed, binary & 19,717 & 44,338 & 3 & No & Yes \\\cline{2-8}
				& Citeseer-M10\footnote{\url{http://citeseerx.ist.psu.edu/}} & directed, binary & 10,310 & 77,218 & 10 & No & Yes \\\hline
				\bf Collaboration network & Arxiv GR-QC \cite{leskovec2007graph} & undirected, binary & 5,242 & 28,980 & N/A & N/A & No \\\hline
				\multirow{3}{*}{\bf Webpage Network} & Wikipedia\footref{fn:Cora} & directed, binary & 2,405 & 17,981 & 20 & No & Yes \\\cline{2-8}
				& WebKB\footref{fn:Cora} & directed, binary & 877 & 1,608 & 5 & No & Yes \\\cline{2-8}
				& Political Blog \cite{adamic2005political} & directed, binary & 1,222 & 16,715 & 2 & No & No \\\hline
				\bf Biological Network & Protein-Protein Interaction \cite{breitkreutz2008biogrid} & undirected, binary & 4,777 & 184,812 & 40 & Yes & No \\\hline
				\bf Communication Network & Enron Email Network\footnote{\url{https://snap.stanford.edu/data/email-Enron.html}} & undirected, binary & 36,692 & 183,831 & 7 & No & No \\\hline
				\bf Traffic Network & European Airline Networks\footnote{\url{http://complex.unizar.es/~atnmultiplex/}} & undirected, binary & N/A & N/A & 4 & No & No \\\hline
			\end{tabular}
	\end{minipage}}
	\label{table-datasets}
\end{table*}

%& Facebook\footnode{\url{https://snap.stanford.edu/data/egonets-Facebook.html}} & undirected, binary & 4,039 & 176,468 & 4 & No & Yes \\\cline{2-8}

Benchmark datasets play an important role for the research community to evaluate the performance of newly developed NRL algorithms as compared to the existing baseline methods. A handful of network datasets have been made publicly available to facilitate the evaluation of NRL algorithms across different tasks. We summarize a list of network datasets used by most of the published network representation learning papers in Table~\ref{table-datasets}.

Table~\ref{table-datasets} summarizes the main characteristics of the publicly available benchmark datasets, including the type of network (directed or undirected, binary or weighted), number of vertices $|V|$, number of edges $|E|$, number of labels $|\mathcal{Y}|$, whether the network is multi-labeled or not, as well as whether network vertices are attached with attributes. In Table~\ref{table-datasets}, according to the property of information networks, we classify benchmark datasets into eight different types:

\noindent\textbf{Social Network}. The BlogCatalog, Flickr and YouTube datasets are formed by users of the corresponding online social network platforms. For the three datasets, vertex labels are defined by user interest groups but user attributes are unavailable. The Facebook network is a combination of 10 Facebook ego-networks, where each vertex contains user profile attributes. The Amherst, Hamilton, Mich and Rochester~\cite{traud2012social} datasets are the Facebook networks formed by users from the corresponding US universities, where each user has six user profile features. Often, user profile features are noisy, incomplete, and long-tail distributed.
	
\noindent\textbf{Language Network}. The language network Wikipedia~\cite{tang2015line} is a word co-occurrence network constructed from the entire set of English Wikipedia pages. There is no class label on this network. The word embeddings learned from this network are evaluated by word analogy and document classification.
	
\noindent\textbf{Citation Network}. The citation networks are directed information networks formed by author-author citation relationships or paper-paper citation relationships. They are collected from different databases of academic papers, such as DBLP and Citeseer. Among the commonly used citation networks, DBLP (AuthorCitation) \cite{tang2015line} is a weighted citation network between authors with the edge weight defined by the number of papers written by one author and cited by the other author, while DBLP (PaperCitation) \cite{tang2015line}, Cora, Citeseer, PubMed and Citeseer-M10 are the binary paper citation networks, which are also attached with vertex text attributes as the content of papers. Compared with user profile features in social networks, the vertex text features here are more topic-centric, informative and can better complement network structure to learn effective vertex representations.
	
\noindent\textbf{Collaboration Network}. The collaboration network Arxiv GR-QC \cite{leskovec2007graph} describes the co-author relationships for papers in the research field of General Relativity and Quantum Cosmology. In this network, vertices represent authors and edges indicate co-author relationships between authors. Because there is no category information for vertices, this network is used for the link prediction task to evaluate the quality of learned vertex representations.
	
\noindent\textbf{Webpage Network}. Webpage networks (Wikipedia, WebKB and Political Blog \cite{adamic2005political}) are composed of real-world webpages and hyperlinks between them, where the vertex represents a webpage and the edge indicates that there is a hyperlink from one webpage to another. Webpage text content is often collected as vertex features.
	
\noindent\textbf{Biological Network}. As a typical biological network, the Protein-Protein Interaction network~\cite{breitkreutz2008biogrid} is a subgraph of the PPI network for Homo Sapiens. The vertex here represents a protein and the edge indicates that there is an interaction between proteins. The labels of vertices are obtained from the hallmark gene sets~\cite{liberzon2011molecular} and represent biological states.
	
\noindent\textbf{Communication Network.} The Enron Email Network is formed by the Email communication between Enron employees, with vertices being employees and edges representing the email communicated between employees. Employees are labeled as 7 roles (\textit{e.g.}, CEO, president and manager), according to their functions.
		
\noindent\textbf{Traffic Network.} European Airline Networks used in ~\cite{donnat2017spectral} are constructed from 6 airlines operating flights between European airports: 4 commercial airlines (Air France, Easyjet, Lufthansa, and RyanAir) and 2 cargo airlines (TAP Portugal, and European Airline Transport). For each airline network, vertices are airports and edges represent the direct flights between airports. In all, 45 airports are labeled as hub airports, regional hubs, commercial hubs, and focus cities, according to their structural roles.

\begin{scriptsize}
	\begin{table*}[h]
		\tabcolsep 5pt
		\renewcommand\arraystretch{1.35}
		\centering
		\caption{A summary of NRL algorithms with respect to the evaluation methodology}
		\scalebox{0.72}{
			\begin{tabular}{|c|c|c|c|c|}
				\hline
				\bf Category &\bf Algorithm & \bf Network Type & \bf Evaluation Method & \bf Code Link\\\hline
				& Social Dim. \cite{tang2009relational,tang2011leveraging,tang2009scalable} & Social Network & Vertex Classification &\\\cline{2-5}
				& DeepWalk \cite{perozzi2014deepwalk} & Social Network & Vertex Classification &\url{https://github.com/phanein/deepwalk}\\\cline{2-5}
				& LINE \cite{tang2015line} &\tabincell{c}{Citation Network\\Language Network\\Social Network} &\tabincell{c}{Vertex Classification\\Visualization} &\url{https://github.com/tangjianpku/LINE}\\\cline{2-5}
				& GraRep \cite{cao2015grarep} & \tabincell{c}{Citation Network\\Language Network\\Social Network} & \tabincell{c}{Vertex Classification\\Vertex Clustering\\Visualization} &\url{https://github.com/ShelsonCao/GraRep}\\\cline{2-5}
				& DNGR \cite{cao2016deep} & Language Network & \tabincell{c}{Vertex Clustering\\Visualization} & \url{https://github.com/ShelsonCao/DNGR}\\\cline{2-5}
				& SDNE \cite{wang2016structural} & \tabincell{c}{Collaboration Network\\Language Network\\Social Network} & \tabincell{c}{Network Reconstruction\\Vertex Classification\\Link Prediction\\Visualization} &\url{https://github.com/suanrong/SDNE}\\\cline{2-5}
				& node2vec \cite{grover2016node2vec} & \tabincell{c}{Biological Network\\Language Network\\Social Network} & \tabincell{c}{Vertex Classification\\Link Prediction} &\url{https://github.com/aditya-grover/node2vec}\\\cline{2-5}
				& HOPE \cite{ou2016asymmetric} & \tabincell{c}{Social Network\\Citation Network} & \tabincell{c}{Network Reconstruction\\Link Prediction} & \\\cline{2-5}
				& APP \cite{zhou2017scalable} & \tabincell{c}{Social Network\\Citation Network\\Collaboration Network} & Link Prediction & \\\cline{2-5}
				& GraphGAN~\cite{wang2018graphgan} & \tabincell{c}{Citation Network\\Language Network\\Social Network} & \tabincell{c}{Vertex Classification\\Link Prediction}& \\\cline{2-5}
				\multirow{-4}{*}{\bf Unsupervised} & M-NMF \cite{wang2017community} & \tabincell{c}{Social Network\\Webpage Network} & \tabincell{c}{Vertex Classification\\Vertex Clustering} & \url{http://git.thumedia.org/embedding/M-NMF} \\\cline{2-5}
				& struct2vec \cite{ribeiro2017struc2vec} & Traffic Network & Vertex Classification & \\\cline{2-5}
				& GraphWave \cite{donnat2017spectral} & \tabincell{c}{Traffic Network\\Communication Network} & \tabincell{c}{Vertex Clustering\\Visualization} & \url{http://snap.stanford.edu/graphwave} \\\cline{2-5}
				& SNS \cite{lyu2017enhancing} & \tabincell{c}{Social Network\\Language Network\\Biological Network} & Vertex Classification & \\\cline{2-5}			
				& DP \cite{feng2018representation} & \tabincell{c}{Social Network\\Citation Network\\Collaboration Network}  & \tabincell{c}{Network Reconstruction\\Link Prediction\\Vertex Classification} & \\\cline{2-5}
				& HARP \cite{chen2018harp} & \tabincell{c}{Social Network\\Collaboration Network\\Citation Network} & \tabincell{c}{Vertex Classification\\Visualization} & \\\cline{2-5}
				& TADW \cite{yang2015network} & \tabincell{c}{Citation Network\\Webpage Network} & \tabincell{c}{Vertex Classification} & \url{https://github.com/thunlp/tadw}\\\cline{2-5}
				& HSCA \cite{zhang2016homophily} & \tabincell{c}{Citation Network\\Webpage Network} & \tabincell{c}{Vertex Classification\\Visualization} & \url{https://github.com/daokunzhang/HSCA} \\\cline{2-5}
				& pRBM \cite{wang2016paired} & Social Network & Vertex Clustering &\\\cline{2-5}
				& UPP-SNE \cite{zhang2017user} & Social Network & \tabincell{c}{Vertex Classification\\Vertex Clustering} &\\\cline{2-5}
				& PPNE \cite{li2017ppne}  & \tabincell{c}{Social Network\\Citation Network\\Webpage network} & \tabincell{c}{Vertex Classification\\Link Prediction} & \\\hline
				& DDRW \cite{li2016discriminative} & Social Network & Vertex Classification &\\\cline{2-5}
				& MMDW \cite{tu2016max} & \tabincell{c}{Citation Network\\ Webpage Network} & \tabincell{c}{Vertex Classification\\Visualization} &\url{https://github.com/thunlp/MMDW}\\\cline{2-5}
				&TLINE \cite{zhang2016tline} & \tabincell{c}{Citation Network\\Collaboration Network} & \tabincell{c}{Vertex Classification\\Visualization} &\\\cline{2-5}
				\multirow{6}{*}{\bf Semi-supervised} & GENE \cite{chen2016incorporate} & Social Network & Vertex Classification & \\\cline{2-5}
				& SemiNE \cite{li2017semi} & Social Network & \tabincell{c}{Network Reconstruction\\Vertex Classification\\Link prediction} & \\\cline{2-5}
				& TriDNR \cite{pan2016tri} & Citation Network & \tabincell{c}{Vertex Classification\\ Visualization} &\url{https://github.com/shiruipan/TriDNR}\\\cline{2-5}
				& LDE \cite{wang2016linked} & \tabincell{c}{Social Network\\Citation Network} & Vertex Classification &\\\cline{2-5}
				& DMF~\cite{zhang2016collective} & Citation Network & \tabincell{c}{Vertex Classification\\Visualization} &\url{https://github.com/daokunzhang/DMF_CC}\\\cline{2-5}
				& Planetoid~\cite{yang2016revisiting} & Citation Network & \tabincell{c}{Vertex Classification\\Visualization} &\url{https://github.com/kimiyoung/planetoid}\\\cline{2-5}
				& LANE \cite{huang2017label} & Social Network & Vertex Classification &\\\hline
		\end{tabular}}
		\label{table-algorithm-summary3}
	\end{table*}
\end{scriptsize}

\subsection{Evaluation Methods}

It is difficult to directly compare the quality of the vertex representations learned by different NRL algorithms, due to the unavailability of ground truth. Alternatively, in order to evaluate the effectiveness of NRL algorithms on learned vertex representations, several network analytic tasks are commonly used for comparison studies.

\noindent\textbf{Network Reconstruction}. The aim of network reconstruction is to reconstruct the original network from the learned vertex representations by predicting the links between vertices based on the inner product or similarity between vertex representations. The known links in the original network serve as the ground truth for evaluating reconstruction performance. $precision@k$ and $MAP$~\cite{wang2016structural} are often used as evaluation metrics. This evaluation method can check whether the learned vertex representations well preserve network structure and support network formation.

\noindent\textbf{Vertex Classification}. As an evaluation method for NRL, vertex classification is conducted by taking learned vertex representations as features to train a classifier on labeled vertices. The classification performance on unlabeled vertices is used to evaluate the quality of the learned vertex representations. Different vertex classification settings, including binary-class classification, multi-class classification, and multi-label classification, are often carried out, depending on the underlying network characteristics. For binary-class classification, $F_{1}$ score is used as the evaluation criterion. For multi-class and multi-label classification, $Micro$-$F_{1}$ and $Macro$-$F_{1}$ are adopted as evaluation criteria.
	
\noindent\textbf{Vertex Clustering}. To validate the effectiveness of NRL algorithms, vertex clustering is also carried out by applying $k$-means clustering algorithm to the learned vertex representations. Communities in networks are served as the ground truth to assess the quality of clustering results, which is measured by $Accuracy$ and $NMI$ (normalized mutual information)~\cite{Strehl03cluster}. The hypothesis is that, if the learned vertex representations are indeed informative, vertex clustering on learned vertex representations should be able to discover community structures. That is, good vertex representations are expected to generate good clustering results.
	
\noindent\textbf{Link Prediction}. Link prediction can be used to evaluate whether the learned vertex representations are informative to support the network evolution mechanism. To perform link prediction on a network, a portion of edges are first removed, and vertex representations are learned from the remaining network. Finally, the removed edges are predicted with the learned vertex representations. The performance of link prediction is measured by $AUC$ and $precision@k$.
	
\noindent\textbf{Visualization}. Visualization provides a straightforward way to visually evaluate the quality of the learned vertex representations. Often, $t$-distributed stochastic neighbor embedding ($t$-SNE)~\cite{maaten2008visualizing} is applied to project the learned vertex representation vectors into a 2-D space, where the distribution of vertex 2-D mappings can be easily visualized. If vertex representations are of good quality, in the 2-D space, vertices within a same class or community should be embedded closely, and the 2-D mappings of vertices in different classes or communities should be far apart from each other.

In Table~\ref{table-algorithm-summary3}, we summarize the type of information networks and network analytic tasks used to evaluate the quality of vertex representations learned by existing NRL algorithms. We also provide hyperlinks for the codes of respective NRL algorithms if available to help interested readers to further study these algorithms or run experiments for comparison. Overall, social networks and citation networks are frequently used as benchmark datasets, and vertex classification is most commonly used as the evaluation method in both unsupervised and semi-supervised settings.

\subsection{Empirical Results}
We observe from the literature that empirical evaluation is often carried out on different datasets under different settings. There is a lack of consistency on empirical results to determine the best performing algorithms and their circumstances. Therefore, we perform benchmark experiments to fairly compare the performance of several representative NRL algorithms on the same set of datasets. Note that, because semi-supervised NRL algorithms are task-dependent: the target task may be binary or multi-class, or multi-label classification, or because they use different classification strategies, it would be difficult to assess the effectiveness of network embedding under the same settings. Therefore, our empirical study focuses on comparing seven unsupervised NRL algorithms (DeepWalk~\cite{perozzi2014deepwalk}, LINE~\cite{tang2015line}, node2vec~\cite{grover2016node2vec}, M-NMF~\cite{wang2017community}, TADW~\cite{yang2015network}, HSCA~\cite{zhang2016homophily}, UPP-SNE~\cite{zhang2017user}) on vertex classification and vertex clustering, which are the two most commonly used evaluation methods in the literature.

\begin{table*}[h]
	\tabcolsep 4pt
	\renewcommand\arraystretch{1.3}
	\caption{Vertex Classification Results on Seven Datasets}
	\centering
	\scalebox{0.65}{	
		\begin{tabular}{cc|ccccccc|ccccccc}
			\hline
			& \multirow{2}{*}{\bf Method}& \multicolumn{7}{c|}{Training ratio = 5\%} & \multicolumn{7}{c}{Training ratio = 50\%} \\\cline{3-16}
			& & Amherst & Hamilton & Mich & Rochester & Citeseer & Cora & Facebook & Amherst & Hamilton & Mich & Rochester & Citeseer & Cora & Facebook \\\hline
			\multirow{7}{*}{$Micro$-$F_{1}$} & DeepWalk & 0.7168 & 0.7127 & 0.3933 & 0.6795 & 0.5061 & 0.7333 & 0.6839 & 0.8106 & 0.8188 & 0.4829 & 0.7822 & 0.5927 & \underline{0.8292} & 0.6782 \\
			& LINE & 0.7351 & 0.7367 & 0.4101 & \textbf{0.7163} & 0.3842 & 0.5625 & 0.6832 & 0.8240 & 0.8415 & \textbf{0.5046} & 0.8067 & 0.5353 & 0.7572 & 0.6848 \\
			& node2vec & \textbf{0.7528} & \textbf{0.7622} & \textbf{0.4163} & 0.7018 & \underline{0.5135} & \underline{0.7395} & \underline{0.6911} & 0.8063 & 0.8239 & 0.4900 & 0.7625 & 0.5936 & 0.8126 & \underline{0.6944} \\
			& M-NMF & 0.7325 & 0.7471 & 0.3865 & 0.7047 & 0.4070 & 0.5704 & 0.6875 & \textbf{0.8280} & \textbf{0.8476} & 0.4827 & \textbf{0.8076} & \underline{0.5979} & 0.7635 & 0.6849\\
			& TADW & & & & & 0.6206 & 0.7257 & 0.7260 & & & & & 0.7379 & 0.8648 & \textbf{0.8748} \\
			& HSCA & & & & & 0.6309 & 0.7737 & 0.6827 & & & & & \textbf{0.7396} & \textbf{0.8693} & 0.6955 \\
			& UPP-SNE & & & & & \textbf{0.6579} & \textbf{0.7745} & \textbf{0.8467} & & & & & 0.7105 & 0.8429 & 0.8711 \\\hline
			\multirow{7}{*}{$Macro$-$F_{1}$} & DeepWalk & 0.3372 & 0.2829 & 0.1726 & 0.1925 & 0.4487 & 0.7103 & \underline{0.2431} & \textbf{0.4628} & \textbf{0.3838} & 0.2249 & \textbf{0.2549} & 0.5281 & \underline{0.8203} & \underline{0.2529} \\
			& LINE & \textbf{0.3420} & 0.2912 & 0.1823 & 0.2043 & 0.3456 & 0.5321 & 0.2350 & 0.4107 & 0.3487 & \textbf{0.2395} & 0.2540 & 0.4851 & 0.7504 & 0.2460 \\
			& node2vec & 0.3158 & 0.2912 & \textbf{0.1825} & 0.1893 & \underline{0.4577} & \underline{0.7193} & 0.2231 & 0.3568 & 0.3211 & 0.2214 & 0.2207 & 0.5370 & 0.8035 & 0.2207 \\
			& M-NMF & 0.3206 & \textbf{0.2951} & 0.1774 & \textbf{0.2050} & 0.3665 & 0.5377 & 0.2183 & 0.3895 & 0.3684 & 0.2341 & 0.2540 & \underline{0.5494} & 0.7554 & 0.2362 \\
			& TADW  & & & & & 0.5614 & 0.7031 & 0.2926 & & & & & \textbf{0.6920} & 0.8527 & \textbf{0.4425} \\
			& HSCA & & & & & 0.5712 & \textbf{0.7544} & 0.2219 & & & & & 0.6909 & \textbf{0.8571} & 0.2459 \\
			& UPP-SNE & & & & & \textbf{0.5847} & 0.7451 & \textbf{0.4177} & & & & & 0.6509 & 0.8277 & 0.4355 \\\hline
	\end{tabular}}
	\label{table-classification-res}
\end{table*}

\begin{table}[h]
	\tabcolsep 4pt
	\renewcommand\arraystretch{1.4}
	\caption{Vertex Clustering Results on on Seven Datasets}
	\centering
	\scalebox{0.65}{
		\begin{tabular}{ccccccccc}
			\hline
			&\bf Method & Amherst & Hamilton & Mich & Rochester & Citeseer & Cora & Facebook \\\hline
			\multirow{7}{*}{$Accuracy$} & DeepWalk & 0.6257 & 0.6273 & 0.3944 & 0.5593 & 0.3365 & 0.5062 & \underline{0.6953} \\
			& LINE & \textbf{0.6908} & \textbf{0.6718} & \textbf{0.4127} & \textbf{0.6070} & 0.2806 & 0.3905 & 0.6952 \\
			& node2vec & 0.6662 & 0.6328 & 0.4114 & 0.5777 & \underline{0.4574} & \underline{0.6216} & 0.6952 \\
			& M-NMF & 0.6545 & 0.6374 & 0.3279 & 0.5071 & 0.2379 & 0.3640 & 0.6952 \\
			& TADW & & & & & 0.2778 & 0.4731 & 0.6953 \\
			& HSCA & & & & & 0.2794 & 0.4594 & 0.6957 \\
			& UPP-SNE & & & & & \textbf{0.5748} & \textbf{0.6832} & \textbf{0.8328} \\\hline
			\multirow{7}{*}{$NMI$} & DeepWalk & 0.4873 & 0.4390 & \textbf{0.1897} & 0.3468 & 0.0896 & 0.3308 & 0.0142 \\
			& LINE & \textbf{0.5030} & \textbf{0.4529} & 0.1858 & \textbf{0.3547} & 0.0511 & 0.1639 & 0.0113 \\
			& node2vec & 0.4742 & 0.4144 & 0.1824 & 0.3193 & \underline{0.2027} & \underline{0.4333} & 0.0162 \\
			& M-NMF & 0.4696 & 0.4330 & 0.1304 & 0.2971 & 0.0464 & 0.1201 & \underline{0.0176} \\
			& TADW  & & & & & 0.0845 & 0.3001 & 0.0651 \\
			& HSCA & & & & & 0.0902 & 0.3148 & 0.0151 \\
			& UPP-SNE & & & & & \textbf{0.3005} & \textbf{0.4911} & \textbf{0.2095} \\\hline
	\end{tabular}}
	\label{table-clustering-res}
\end{table}

Our empirical studies are based on seven benchmark datasets: Amherst, Hamilton, Mich, Rochester, Citeseer, Cora and Facebook. Following ~\cite{wang2017community}, for Amherst, Hamilton, Mich and Rochester, only the network structure is used and the attribute ``year'' is used as class label, which is a good indicator of community structure. For Citeseer and Cora, the research area is used as the class label. The class label of Facebook dataset is given by the attribute ``education type''.

\subsubsection{Experimental Setup}

For random walk based methods, DeepWalk, node2vec and UPP-SNE, we uniformly set the number of walks, walk length and window size as 10, 80, 10, respectively.  For UPP-SNE, we use the implementation that is optimized by stochastic gradient descent. The parameter $p$ and $q$ of node2vec are set to 1, as the default setting. For M-NMF, we set $\alpha$ and $\beta$ as 1. For all algorithms, the dimension of learned vertex representations is set to 256. For LINE, we learn 128-dimensional vertex representations with the first-order proximity preserving version and the second-order proximity preserving version respectively and concatenate them together to obtain 256-dimensional vertex representations. The other parameters of the above algorithms are all set to their default values.

Taking the learned vertex representations as input, we carry out vertex classification and vertex clustering experiments to evaluate the quality of learned vertex representations. For vertex classification, we randomly select 5\% and 50\% samples to train an SVM classifier (with the LIBLINEAR implementation~\cite{fan2008liblinear}) and test it on the remaining samples. We repeat this process 10 times and report the averaged $Micro$-$F_{1}$ and $Macro$-$F_{1}$ values. We adopt $K$-means to perform vertex clustering. To reduce the variance caused by random initialization, we repeat the clustering process for 20 times and report the averaged $Accuracy$ and $NMI$ values.

\subsubsection{Performance Comparison}
Table~\ref{table-classification-res} and \ref{table-clustering-res} compare the performance of different algorithms on vertex classification and vertex clustering.  For each dataset, the best performing method across all baselines is \textbf{bold-faced}. For the attributed networks (Citeseer, Cora and Facebook), the underlined results indicate the best performer among the structure only preserving NRL algorithms (DeepWalk, LINE, node2vec and M-NMF). 

Table \ref{table-classification-res} shows that among structure only preserving NRL algorithms, when the training ratio is 5\%, node2vec achieves the best classification performance overall, and when the training ratio is 50\%, M-NMF performs best in terms of $Micro$-$F_{1}$ while DeepWalk is the winner of $Macro$-$F_{1}$. Here, M-NMF does not exhibit significant advantage over DeepWalk, LINE and node2vec. This is probably due to that the parameter $\alpha$ and $\beta$ of N-NMF are not optimally tuned; their values must be carefully chosen so as to achieve a good trade-off between different components. On attributed networks (Citeseer, Cora and Facebook), the content augmented NRL performs much better than the structure only preserving NRL algorithms. This proves that vertex attributes can largely contribute to learning more informative vertex representations. When training ratio is 5\%, UPP-SNE is the best performer. This indicates that the UPP-SNE's non-linear mapping provides a better way to construct vertex representations from vertex attributes than the linear mapping, as is done in TADW and HSCA. When training ratio is 50\%, TADW achieves the best overall classification performance, although in some cases, it is slightly outperformed by HSCA. On citation networks (Citeseer and Cora), HSCA performs better than TADW, while it yields worse performance than TADW on Facebook. This might be caused by the fact that the homophily property of Facebook social network is weaker than that of citation networks. The homophily preserving objective should be weighted less to make HSCA achieve satisfactory performance on Facebook. 

%{\color{blue}Form Table \ref{table-classification-res}, we observe that on Amherst, Hamilton, Mich, Rochester, the only structure preserving NRL algorithms (DeepWalk, LINE, node2vec and M-NMF) achieves similar classification performance, with an exception in the $Macro$-$F_{1}$ values on the Amherst dataset. Though M-NMF performs slightly better than DeepWalk, LINE and node2vec in terms of $Micaro$-$F_{1}$, it does not present significant advantage over them. Maybe the parameter $\alpha$ and $\beta$ of N-NMF need to be carefully turned to make different components better trade off with each other so as to achieve good performance. On the attributes information networks (Citeseer, Cora and Facebook), the content augmented NRL algorithms achieve the best classification performance, with the winner UPP-SNE on Citeseer and Cora, and the winner HSCA on PubMed. This proves that vertex content attributes can be effectively leveraged by these content augmented NRL algorithms to learn more informative vertex representations. }

\begin{table*}[tb]
	\centering
	\scriptsize
	\tabcolsep 5pt
	\renewcommand\arraystretch{1.4}
	\caption{Complexity analysis}
	\scalebox{0.85}{
	\begin{tabular}{|c|c|c|c|}
		\hline
		{\bf Category} & {\bf Algorithm} & {\bf Complexity} & {\bf Optimization Method}\\
		\hline
		\multirow{18}{*}{\bf Unsupervised}&Social Dim. \cite{tang2009relational,tang2011leveraging,tang2009scalable} & $O(d|V|^{2})$ & \multirow{4}{*}{\bf Eigen Decomposition} \\\cline{2-3}
		&GraRep \cite{cao2015grarep} & $O(|V||E|+d|V|^{2})$ & \\\cline{2-3}
		&HOPE \cite{ou2016asymmetric} & $O(d^{2}I|E|)$ & \\\cline{2-3}
		&GraphWave \cite{donnat2017spectral} & $O(|E|)$ & \\\cline{2-4}
		&DeepWalk \cite{perozzi2014deepwalk} & $O(d|V|\log|V|)$ & \multirow{10}{*}{\bf Stochastic Gradient Descent} \\\cline{2-3}
		&LINE \cite{tang2015line} & $O(d|E|)$ & \\\cline{2-3}
		&SDNE \cite{wang2016structural} & $O(dI|V|^{2})$ & \\\cline{2-3}
		&node2vec \cite{grover2016node2vec} & $O(d|V|)$ & \\\cline{2-3}
		&APP \cite{zhou2017scalable} & $O(d|V|)$ & \\\cline{2-3}
		&GraphGAN \cite{wang2018graphgan} & $O(|V|\log|V|)$ &\\\cline{2-3}
		&struct2vec \cite{ribeiro2017struc2vec} & $O(|V|^{3})$ &\\\cline{2-3}
		&SNS \cite{lyu2017enhancing} & $O(d|V|)$ &\\\cline{2-3}
		&pRBM \cite{wang2016paired} & $O(dmI|V|)$ & \\\cline{2-3}
		&PPNE \cite{li2017ppne} & $O(d|V|)$ &\\\cline{2-4}
		&M-NMF \cite{wang2017community} & $O(dI|V|^{2})$ & \multirow{3}{*}{\bf Alternative Optimization} \\\cline{2-3}
		&TADW \cite{yang2015network} & $O(|V||E|+dI|E|+dmI|V|+d^{2}I|V|)$ &\\\cline{2-3}
		&HSCA \cite{zhang2016collective} & $O(|V||E|+dI|E|+dmI|V|+d^{2}I|V|)$ &\\\cline{2-4}
		&UPP-SNE \cite{zhang2017user} & $O(I|E|\cdot nnz(X))$ & \bf Gradient Descent \\\cline{1-4}
		\multirow{7}{*}{\bf Semi-supervised}& DDRW \cite{li2016discriminative} & $O(d|V|\log|V|)$ & \multirow{5}{*}{\bf Stochastic Gradient Descent} \\\cline{2-3}
		&TLINE \cite{zhang2016tline} & $O(d|E|)$ & \\\cline{2-3}
		&SemiNE \cite{li2017semi} & $O(d|V|)$ & \\\cline{2-3}
		&TriDNR \cite{pan2016tri} & $O(d\cdot nnz(X)\log m+d|V|\log|V|)$ &\\\cline{2-3}
		&LDE \cite{wang2016linked}  & $O(dI\cdot nnz(X)+dI|E|+dI|\mathcal{Y}||V|)$ &\\\cline{2-4}
		&DMF \cite{zhang2016collective} & $O(|V||E|+dI|E|+dmI|V|+d^{2}I|V|)$ & \multirow{2}{*}{\bf Alternative Optimization} \\\cline{2-3}
		&LANE \cite{huang2017label} & $O(m|V|^{2}+dI|V|^{2})$ &\\\hline
	\end{tabular}}
	\label{table-complexity}
\end{table*}

Table~\ref{table-clustering-res} shows that LINE achieves the best clustering performance on Amherst, Hamilton, Mich and Rochester. As LINE's vertex representations capture both the first-order and second-order proximity, it can better preserve the community structure, leading to good clustering performance. On Citeseer, Cora and Facebook, the content augmented NRL algorithm UPP-SNE performs best. As UPP-SNE constructs vertex representations from vertex attributes via a non-linear mapping, the well preserved content information favors the best clustering performance. On Citeseer and Cora, node2vec performs much better than other structure only preserving NRL algorithms, including its equivalent version DeepWalk. For each vertex context pair $(v_{i},v_{j})$, DeepWalk and node2vec use two different strategies to approximate the probability $\mathrm{Pr}(v_{j}|v_{i})$: hierarchical softmax~\cite{mnih2009scalable,morin2005hierarchical} and negative sampling~\cite{gutmann2012noise}. The better clustering performance of node2vec over DeepWalk proves the advantage of negative sampling over hierarchical softmax, which is consistent with the word embedding results as reported in~\cite{le2014distributed}.

\subsection{Complexity Analysis}

To better understand the existing NRL algorithms, we provide a detailed analysis of their time complexity and underlying optimization methods in Table~\ref{table-complexity}. A new notation $I$ is introduced to represent the number of iterations and we use $nnz(\cdot)$ to denote the number of non-zero entries of a matrix. In a nutshell, four kinds of solutions are used to optimize the objectives of the existing NRL algorithms: (1) \textbf{eigen decomposition} that involves finding top-$d$ eigenvectors of a matrix, (2) \textbf{alternative optimization} that optimizes one variable with the remaining variables fixed alternately, (3) \textbf{gradient descent} that updates all parameters at each iteration for optimizing the overall objective, and (4) \textbf{stochastic gradient descent} that optimizes the partial objective stochastically in an on-line mode. 

Both unsupervised and semi-supervised NRL algorithms mainly adopt stochastic gradient descent to solve their optimization problems. The time complexity of these algorithms is often linear with respect to the number of vertices/edges, which makes them scalable to large-scale networks. By contrast, other optimization strategies usually involve higher time complexity, which is quadratic with regards to the number of vertices, or even higher with the scale of the number of vertices times the number of edges. The corresponding NRL algorithms usually perform factorization on a $|V|\times|V|$ structure preserving matrix, which is quite time-consuming. Efforts have been made to reduce the complexity of matrix factorization. For example, TADW~\cite{yang2015network}, DMF~\cite{zhang2016collective} and HSCA~\cite{zhang2016homophily} take advantage of the sparsity of the original vertex-context matrix. HOPE~\cite{ou2016asymmetric} and GraphWave~\cite{donnat2017spectral} adopt advanced techniques~\cite{hochstenbach2009jacobi}~\cite{shuman2011chebyshev} to perform matrix eigen decomposition.
%Here, the time complexity of MMDW, GENE, Planetoid and DNGR is not given due to the lack of technique details in the original papers.

%The time complexity comparison in Table~\ref{table-methodology} shows that matrix factorization based approaches usually require quadratic time complexity with respect to number of vertices. This prevents them from scaling to large datasets. Random walk, edge modeling, and deep learning based methods that mainly adopt stochastic gradient descent optimization strategy are much more efficient than matrix factorization based methods that are solved by eigen decomposition and alternative optimization.

\section{Future Research Directions}
\label{sec:research-direction}

In this section, we summarize six potential research directions and future challenges to stimulate research on network representation learning.

\noindent\textbf{Task-dependence:} To date, most existing NRL algorithms are task-independent, and task-specific NRL algorithms have primarily focused on vertex classification under the semi-supervised setting. Only very recently, a few studies have started to design task-specific NRL algorithms for link prediction~\cite{ou2016asymmetric}, community detection~\cite{cavallari2017learning,yang2018multi,zhang2018cosine,wang2017mgae}, class imbalance learning~\cite{wang2018rsdne}, active learning~\cite{huang2018exploring}, and information retrieval~\cite{Misra2018bernoulli}. The advantage of using network representation learning as an intermediate layer to solve the target task is that the best possible information preserved in the new representation can further benefit the subsequent task. Thus, a desirable task-specific NRL algorithm must preserve information critical to the specific task in order to optimize its performance.  

%At the core of task-specific network representation learning is the ability to identify and preserve the desired network properties that can best benefit the target task.

%For example, semi-supervised NRL algorithms leverage vertex labels to learn discriminative vertex representations for maximizing vertex classification performance. Algorithms like HOPE \cite{ou2016asymmetric} and APP \cite{song2009scalable} learn effective vertex representations for link prediction by preserving the asymmetric structural proximity. {\color{blue}Up to now, some network representation learning algorithms have been proposed for the task of community detection~\cite{cavallari2017learning,yang2018multi,zhang2018cosine,wang2017mgae}, class imbalance classification~\cite{wang2018rsdne}, active learning~\cite{huang2018exploring}, and information retrieval~\cite{Misra2018bernoulli}.}

\noindent\textbf{Theory:} Although the effectiveness of the existing NRL algorithms has been empirically proved through experiments, the underlying working mechanism has not been well understood. There is a lack of theoretical analysis with regard to properties of algorithms and what contributes to good empirical results. To better understand DeepWalk~\cite{perozzi2014deepwalk}, LINE~\cite{tang2015line}, and node2vec~\cite{grover2016node2vec}, ~\cite{qiu2018network} discovers their theoretical connections to graph Laplacians. However, in-depth theoretical analysis about network representation learning is necessary, as it provides a deep understanding of algorithms and helps interpret empirical results.

%Theoretical analysis about algorithm properties not only help interpret the empirical results, but also help people better understand the algorithm and facilitate its practical application. To better understand DeepWalk~\cite{perozzi2014deepwalk}, LINE~\cite{tang2015line}, and node2vec~\cite{grover2016node2vec}, ~\cite{qiu2018network} discovers the theoretical connections between them and graph Laplacians. However, more in-depth theoretical analysis about network representation learning algorithm is necessary.

\noindent\textbf{Dynamics:} Current research on network representation learning has mainly concerned static networks. However, in real-life scenarios, networks are not always static. The underlying network structure may evolve over time, i.e., new vertices/edges appear while some old vertices/edges disappear. The vertices/edges may also be described by some time-varying information. Dynamic networks have unique characteristics that make static network embedding fail to work: (i) vertex content features may drift over time; (ii) the addition of new vertices/edges requires learning or updating vertex representations to be efficient; and (iii) network size is not fixed. The work on dynamic network embedding is rather limited; the majority of existing approaches (\textit{e.g.}, \cite{yang2017properties,li2017attributed,zhou2018dynamic}) assume that the node set is fixed and deal with the dynamics caused by the deletion/addition of edges only. However, a more challenging problem is to predict the representations of new added vertices, which is referred to as ``out-of-sample" problem. A few attempts such as~\cite{yang2016revisiting,hamilton2017inductive,ma2018depthlgp} are made to exploit inductive learning to address this issue. They learn an explicit mapping function from a network at a snapshot, and use this function to infer the representations of out-of-sample vertices, based on their available information such as attributes or neighborhood structure. However, they have not considered how to incrementally update the existing mapping function. How to design effective and efficient representation learning algorithms in complex dynamic domains still requires further exploration. 

\noindent\textbf{Scalability:} The scalability is another driving factor to advance the research on network representation learning. Several NRL algorithms have made attempts to scale up to large-scale networks with linear time complexity with respect to the number of vertices/edges. Nevertheless, the scalability still remains a major challenge. Our findings on complexity analysis show that random walk and edge modeling based methods that adopt stochastic gradient descent optimization are much more efficient than matrix factorization based methods that are solved by eigen decomposition and alternative optimization. Matrix factorization based methods have shown great promise in incorporating vertex attributes and discovering community structures, but their scalability needs to be improved to handle networks with millions or billions of vertices. Deep learning based methods can capture non-linearity in networks, but their computational cost is usually high. Traditional deep learning architectures take advantage of GPU to speed up training on Euclidean structured data~\cite{rossi2017deep}. However, networks do not have such a structure, and therefore require new solutions to improve the scalability~\cite{bronstein2017geometric}. 

%The high dimensionality of vertex attributes also incurs extra overhead for network representation learning. 

%like on-line matrix factorization or parallel matrix factorization 

%\textbf{Non-linearity.} Earlier research has primarily focused on preserving various types of structure information and demonstrated great promise in various tasks. More recent studies have devoted to leveraging the information from vertex attributes and/or vertex labels to enhance network representation learning. From the methodology perspective, the majority of the existing techniques are matrix factorization and random walk based, while deep learning based methods have explored the use of non-linear models to capture more complex network structure. Yet, finding effective ways to improve the interpretability of deep learning based methods remains an open research problem.

\noindent\textbf{Heterogeneity and semantics:} Representation learning for heterogeneous information networks (HIN) is one promising research direction. The vast amounts of existing work has focused on homogeneous network embedding, where all vertices are of the same type and edges represent a single relation. However, there is an increasing need to study heterogeneous information networks with different types of vertices and edges, such as DBLP, DBpedia, and Flickr. An HIN is composed of different types of entities, such as text, images, or videos, and the interdependencies between entities are very complex. This makes it very difficult to measure rich semantics and proximity between vertices and seek a common and coherent embedding space. Recent studies by~\cite{chang2015heterogeneous,huang2017heterogeneous,Dong:2017:metapath2vec,liu2018distance,tu2017structural,ma2018multi,zhang2017learning,qu2017attention,fu2017hin2vec,chen2017hine} have investigated the use of various descriptors (\textit{e.g.}, metapath or meta structure) to capture semantic proximity between distant HIN vertices for representation learning. However, the research along this line is still at early stage. Further research requires to investigate better ways for capturing the proximity between cross-modal data, and their interplay with network structure.

Another interesting direction is to investigate edge semantics in signed networks, where vertices have both positive and negative relationships. Signed networks are ubiquitous in social networks, such as Epinions and Slashdot, that allow users to form positive or negative friendship/trust connection to other users. The existence of negative links makes the traditional homophily based network representation learning algorithms unable to be directly applied. Some studies~\cite{wang2017signed,wang2017attributed,wang2017shine} tackle signed network representation learning through directly modeling the polar of links. How to fully encode network structure and vertex attributes for signed network embedding remains an open question.

%{\color{blue}
%\textbf{Signed networks}.  Signed network is a special type of information network, which has both positive and negative relationships. Signed networks are ubiquitous in social media in forms of trust networks, friend-foe networks, and sentiment networks, \textit{etc.} The existence of negative links makes the traditional homophily based network representation learning algorithms unable to be directly applied. How to incorporate the polar of links and preserve network structure and vertex attributes for signed network representation is a challenging research problem. Some efforts~\cite{wang2017signed,wang2017attributed,wang2017shine} have been made for signed network representation learning. However, they are all achieved by directly modeling the polar of links. Future work should be carried out for capturing more global structure.
%}

\noindent\textbf{Robustness:} Real-world networks are often noisy and uncertain, which makes traditional NRL algorithms unable to produce stable and robust representations. ANE (Adversarial Network Embedding)~\cite{dai2018adversarial} and ARGA (Adversarially
Regularized Graph Autoencoder)~\cite{pan2018adversarially} learn robust vertex representations via enforcing an adversarial learning regularizer~\cite{goodfellow2014generative}. To deal with the uncertainty in the existence of edges, URGE (Uncertain Graph Embedding)~\cite{hu2017embedding} encodes the edge existence probability into the vertex representation learning process. It is of great importance to have more research efforts on enhancing the robustness of network representation learning.  

\section{Conclusion}
\label{sec:conclusion}

This survey provides a comprehensive review of the state-of-the-art network representation learning algorithms in the data mining and machine learning field. We propose a taxonomy to summarize existing techniques into two settings: unsupervised setting and semi-supervised settings. According to the information sources they use and the methodologies they employ, we further categorize different methods at each setting into subgroups, review representative algorithms in each subgroup, and compare their advantages and disadvantages. We summarize evaluation protocols used for validating existing NRL algorithms, compare their empirical performance and complexity, as well as point out a few emerging research directions and the promising extensions. Our categorization and analysis not only help researchers to gain a comprehensive understanding of existing methods in the field, but also provide rich resources to advance the research on network representation learning.

\section*{Acknowledgments}
The work was supported by the US National Science Foundation (NSF) through grant IIS-1763452, and the Australian Research Council (ARC) through grant LP160100630 and DP180100966. Daokun Zhang was supported by China Scholarship Council (CSC) with No. 201506300082 and a post-graduate scholarship from Data61, CSIRO in Australia.

\bibliographystyle{IEEEtran}
\bibliography{NRL-bibliography}
\end{document}